\documentclass[sigconf]{acmart}
\usepackage{color}
\usepackage{booktabs} 
\usepackage{subfig}
\usepackage{amssymb}
\usepackage{amsmath}
\usepackage{graphicx}

\graphicspath{ {./figure/} }

\setcopyright{none}

\begin{document}
\title{What did you see?\\ Personalization, regionalization and the question of the filter bubble in Google's search engine}

\author{Tobias D. Krafft}
\orcid{0000-0002-3527-1092}
\affiliation{%
  \institution{Algorithm Accountablitiy Lab}
  \streetaddress{Gottlieb-Daimler-Str. 48}
  \city{TU Kaiserslautern}\\
  \includegraphics[scale=0.5]{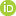} \href{https://orcid.org/0000-0002-3527-1092}{0000-0002-3527-1092}\\
	}
\email{krafft@cs.uni-kl.de}

\author{Michael Gamer}
\orcid{0000-0003-0261-0921}
\affiliation{%
  \institution{Algorithm Accountablitiy Lab}
  \streetaddress{Gottlieb-Daimler-Str. 48}
  \city{TU Kaiserslautern}\\
  \includegraphics[scale=0.5]{orcid_16x16} \href{ https://orcid.org/0000-0003-0261-0921}{0000-0003-0261-0921}\\
	}
\email{gamer@cs.uni-kl.de}

\author{Katharina A. Zweig}
\orcid{0000-0002-4294-9017}
\affiliation{%
  \institution{Algorithm Accountablitiy Lab}
  \streetaddress{Gottlieb-Daimler-Str. 48}
  \city{TU Kaiserslautern}\\
	\includegraphics[scale=0.5]{orcid_16x16} \href{https://orcid.org/0000-0002-4294-9017}{0000-0002-4294-9017}\\
	}
\email{zweig@cs.uni-kl.de}

\renewcommand{\shortauthors}{Krafft, Gamer and Zweig}

\begin{abstract}
This report analyzes the Google search results from more than 1,500 volunteer data donors who, in the five weeks leading up to the federal election on September 24th, 2017, automatically searched Google for 16 predefined names of political parties and politicians every four hours. It is based on an adjusted database consisting of more than 8,000,000 data records, which were generated in the context of the research project "\#Datenspende: Google und die Bundestagswahl 2017" and sent to us for evaluation. The \#Datenspende project was commissioned by six state media authorities. Spiegel Online acted as a media partner. Focal points of the present study are i. a. the question of the degree of personalization of search results, the proportion of regionalization and the risk of algorithm-based filter bubble formation or reinforcement by the leader in the search engine market.
\end{abstract}


\maketitle

\section{Introduction: The filter bubble's dangers to society}
Political formation of opinion as well as general access to information has changed significantly due to digitalization. Information sources such as newspapers, TV and radio, where large parts of the population read, heard, or saw the same news and interpretations of these news stories, are being replaced by more and more diverse and personalized media offerings as a result of digital transformation. Personalization is enabled by algorithm-based systems: here, an algorithm -- not a human -- decides which contents users might be interested in, and only these are offered to them in social networks. The same is true for personalized search engines like Google, Yahoo or Bing, for which, at more than 3,200 billion search queries in 2016 (Statista, 2016), it is impossible to have a man-made sorting available.
\subsection{The model of algorithmically generated and amplified filter bubbles}
The possibilities and dangers of a so-called algorithmically generated filter bubble increase steadily. This term is to be understood as a partial concept of the filter bubble theory by Eli Pariser. In his 2011 book, "The Filter Bubble: What the Internet Is Hiding from You" (Pariser, 2011), the internet activist pointed out the possible dangers of so-called filter bubbles. In his TED talk and on the basis of two screenshots from 2011, he showed that two of his friends had received significantly different results when searching for "Egypt" on the online search platform Google\footnote{See Eli Pariser's TED talk "Beware the filter bubbles", \href{https://www.ted.com/talks/eli_pariser_beware_online_filter_bubbles}{https://www.ted.com/talks/eli\_pariser\_beware\_online\_filter\_bubbles} (accessed May 12th 2018)}. From this he developed a theory according to which personalized algorithms in social media tend to present individuals with content that matches the user's previous views, leading to the emergence of different information spheres where different contents or opinions prevail.\\
In short, filtering the information flow individually can result in groups or individuals being offered different facts, thus living in a unique informational universe\footnote{"a unique universe of information for each of us" (Pariser, 2011, p. 9).}. This is especially worrying if the respective content is politically extreme in nature and if a one-sided perspective results in impairment or total deterioration of citizens' discursive capabilities. A \textbf{filter bubble} in this sense is a selection of news that corresponds to one's own perspectives, which could potentially lead to solidification of one's own position in the political sphere\footnote{"Filter bubbles" can be understood as even more advanced concepts. Selecting websites by means of state cencorship can also create filter bubbles by restricting information. While this cencorship is supported by algorithms, this does not constitute a filter bubble through algorithm-based personalization that is hypothesized in this text. This kind of restriction of information and its possible consequences for filter bubble formation aren't examined here. After all, a search engine operator or a social media platform could knowingly and intentionally limit the data base in a certain direction, thus presenting all users with the same content while only offering a selective extract of reality. This option is not examined here.}.
\subsubsection{Filtering search results with personalization and regionalization}
Since the number of web pages associated with a search term is more than 10 for most queries and at the same time the first 10 web pages shown receive the greatest attention from users, it is essential that search engines filter the possible search results. Certainly one of the most important \textbf{filters} is the user's language, while topicality and popularity play an additional role as well as, to a lesser extent, embedment in the entire WWW (e.g. measured by the PageRank).\\
A particularly important filtering mechanism within the framework of Eli Pariser's filter bubble theory is "personalization". Regarding the term of (preselected) \textbf{personalization}, we follow the statements given by Zuidserveen Borgesius et al. (2016), according to which personalization allows the selection of content that has not yet been clicked by the user, but which is associated with users with similar interests. Algorithmically speaking, this is based on so-called "recommendation systems", which determine the interests of a currently searching user from other people who have shown similar click behavior in the past. It is also also plausible that according to their own click behavior and together with known categorizations of clicked content a profile is compiled for each person, saying, for instance: "This person prefers news about sports and business, reads medium-length text and news that are not older than a day." (Weare, 2009). Considering the vast number of users, both methods can only be achieved algorithmically, using different modes of machine learning and thus only form statistical models (Zweig et al., 2017). Generally it can be expected that users logged into their Google accounts will tend to receive more personalized search results.\\
Websites from the search result list which have previously been clicked by a user are delimited from personalization. For outsiders, those might not be differentiable from the entries that have been personalized by algorithms, but regarding content they don't contribute to algorithm-based filter bubble formation or amplification, since users have chosen those contents before.\\
We use the term \textbf{regionalization} for a sample of websites for a whole group of persons who currently search from a specific region or are known to originate from a specific region, yet don't necessarily mention a region in their search request. For instance, the current location can roughly be derived from the searching device's IP address, or more accurately from smartphone location information or from the profile known to the search engine (Teevan et al., 2011). The delivered websites themselves clearly relate to the location of interest specified by Google; which can be the case, for example, if a nearby location's name appears on the website repeatedly.\\
It is important to note that regionalization on a particularly small scale can be counted towards personalization -- for example, if a selection of regional websites is delivered to each person of a household while differing from the selection for their neighbors.
\subsubsection{When are algorithmically generated and amplified filter bubbles dangerous?}
Eli Pariser's filter bubble theory, with its unsettling consequences for society, is based on these four basic mechanisms:
\begin{enumerate}
  \item{\textbf{Personalization}: An individually customized selection of contents, which achieves a new level of granularity and previously unknown scalability.}
	\item{\textbf{Low overlap of respective news results}: A low or non-existent overlap of filter bubbles, i. e. news and information from one group remain unknown to another.}
	\item{\textbf{Contents}: The nature of the content, which essentially only becomes problematic with politically charged topics and drastically different perspectives.}
	\item{\textbf{Isolation from other sources of information}: The groups of people whose respective news situation displays homogeneous, politically charged and one-sided perspectives, rarely use other sources of information or only those which place them in extremely similar filter bubbles.}
\end{enumerate}
The stronger those four mechanisms manifest themselves, the stronger the filter bubble effect grows, including its harmful consequences for society. The degree of personalization is essential, as politically relevant filter bubbles do not emerge if personalization of an algorithm responsible for selecting news is low. Higher or high personalization and verifiable filter bubbles do not necessarily take political effect if either their contents aren't political in nature or users make use of other sources of information as well. For instance, information delivered to citizens of different languages are free of overlap by definition, if the results are displayed in those languages -- regardless, content-wise those citizens aren't in any way embedded in filter bubbles.
\subsection{Examining Eli Pariser's filter bubble theory}
As algorithms are capable of controlling the flow of information directed towards users, they are assigned a gatekeeper role similar to journalists in traditional journalism (see Moe \& Syvertsen, 2007). As a result, it is necessary to examine how powerful the algorithmically generated and hardened filter bubbles on various intermediaries and search engines actually are. The number of reliable studies is relatively low: an important German study by the Hans Bredow institute offers a positive answer to the question of the informational mix: sources of information today are diverse and capable of pervading other news and information of algorithmically generated and hardened filter bubbles (Schmidt et al., 2017). It is pointed out that algorithms offer a possibility to burst open filter bubbles if such a functionality is explicitly implemented.\\
To our knowledge, apart from anecdotal examinations, a quantitative evaluation of the degree of personalization for a larger user base hasn't been deducted up until 2017: for example, in the context of a Slate article Jacob Weisberg asked five persons to search for topics and found results to be very similar (Weisberg, 2011).\\
Vital questions of the degree of personalization and overlap of single news flows can only be resolved with a large user base. Executing such an investigation appears imperative, especially in light of the debate regarding influence of filter bubbles in social networks, which was sparked in 2016 after Donald Trump's presidential election victory -- unfortunately, due to insufficient APIs\footnote{An API is an interface which enables automated interaction with a program. For example, information can directly be requested from data bases by use of an API.}, this is currently not possible. Given the major political event of the federal elections we decided to realize the \#Datenspende: project in order to find out whether Google already personalizes search results, as has often been speculated. Additionally, it is much simpler to carry out such an examination on search engines, whose results are presented on an HTML page that can easily be processed further. With this project we have introduced a study design that is capable of answering this question automatically for any sample of search engine users and for any search request. Owing to its design, users were able to comprehend at all times which search requests we presented to Google by using their account. As a result, the data collection was trustworthy and the required software was downloaded more than 4000 times.\\
Furthermore, the study design represents a proof of concept, as it enables society to permanently monitor search engines' degree of personalization for any desired search terms. The general design can also be transferred to intermediaries, if appropriate APIs restrict selective access to content relevant to the study in order to establish a similar degree of trustworthiness. For example, on Facebook this would mean selective access to media messages or political advertisements in a respective time-line, while excluding access to private messages from friends.

\section{Study design}
The study design as well as the fundamentals of the survey are explained below. This includes the data structure, relevant terminology and processing of the data basis.
\subsection{Software structure and enrollment}

\begin{figure}[ht!]
    \centering
    \includegraphics[width=0.5\textwidth]{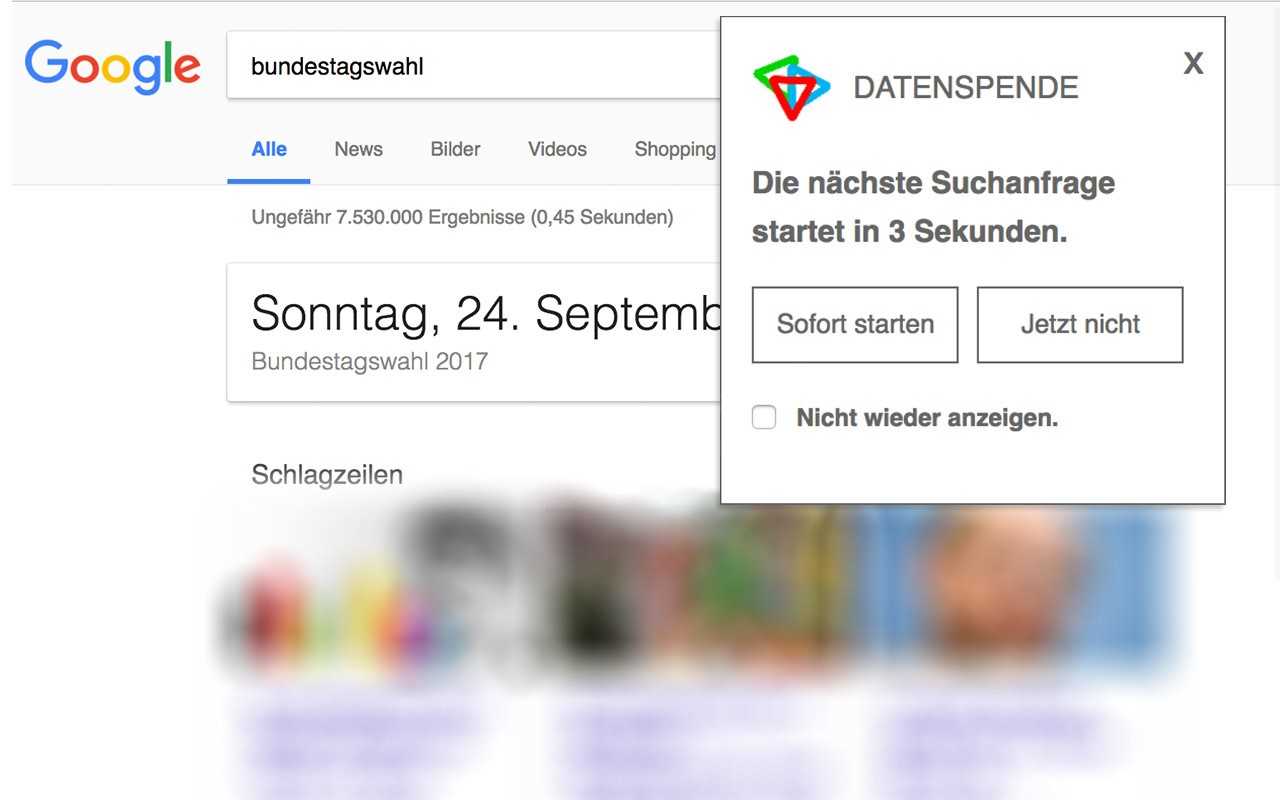}
    \caption{Browser plug-in, immediately before initiating a search for a user whose first result page then was transfered (and thus "donated") to the provided server structure.}
\end{figure}

The basic tool used for data collection is a plug-in that is easily integrated into the web browser and then utilizes the donating user's browser to automatically run search requests and send the data to a central server. The plug-in was created in cooperation with the company ‚lokaler' and was available for the web browsers Chrome and Firefox in order to achieve a market coverage of more than 60\% in Germany (Statista, 2018). All necessary insights into the plug-in source code were released at the start of the project\footnote{https://github.com/algorithmwatch/datenspende, published June 7th 2017.}. At fixed points in time (4:00, 8:00, 12:00, 16:00, 20:00 and 0:00), the plug-in searched for 16 search terms, given that the browser was open at that time\footnote{If the browser was turned off at one or multiple times of search, a cycle of search requests would be started when turning it on again the next time -- which is why there are search result lists with different time stamps. Additionally, it was possible to manually start a search cycle in one's own browser.}. The search requests to Google and Google News proceeded automatically and the donors' personal results were automatically sent to our data donation servers. Consequently, for each user and point in time 16 search terms were requested twice and the respective first page of the search results was saved.\\
The search terms are limited to the seven major political parties and their respective leaders (see Table 1). As can be seen in Figure 1, after downloading the plug-in users were free to decide whether they wanted to be informed about future donations or if those should predominantly run in the background.\\
Information regarding the project and the related call for the data donation were distributed via our project partners' communication channels as well as our media partner Spiegel Online (Horchert, 2017). As a result, 4,384 plug-in installations took place. The resulting search results are freely accessible to the public for analytical purposes\footnote{https://datenspende.algorithmwatch.org/data.html}.\\
It should be pointed out that all results that can be seen in the final report are not necessarily representative, as the data donors were recruited voluntarily and by self-selection. For the most vital findings however, especially regarding the degree of personalization, we assume that they do not change much if the user base is representative.\\

\begin{table}[ht!]
    \centering
    \includegraphics[width=0.5\textwidth]{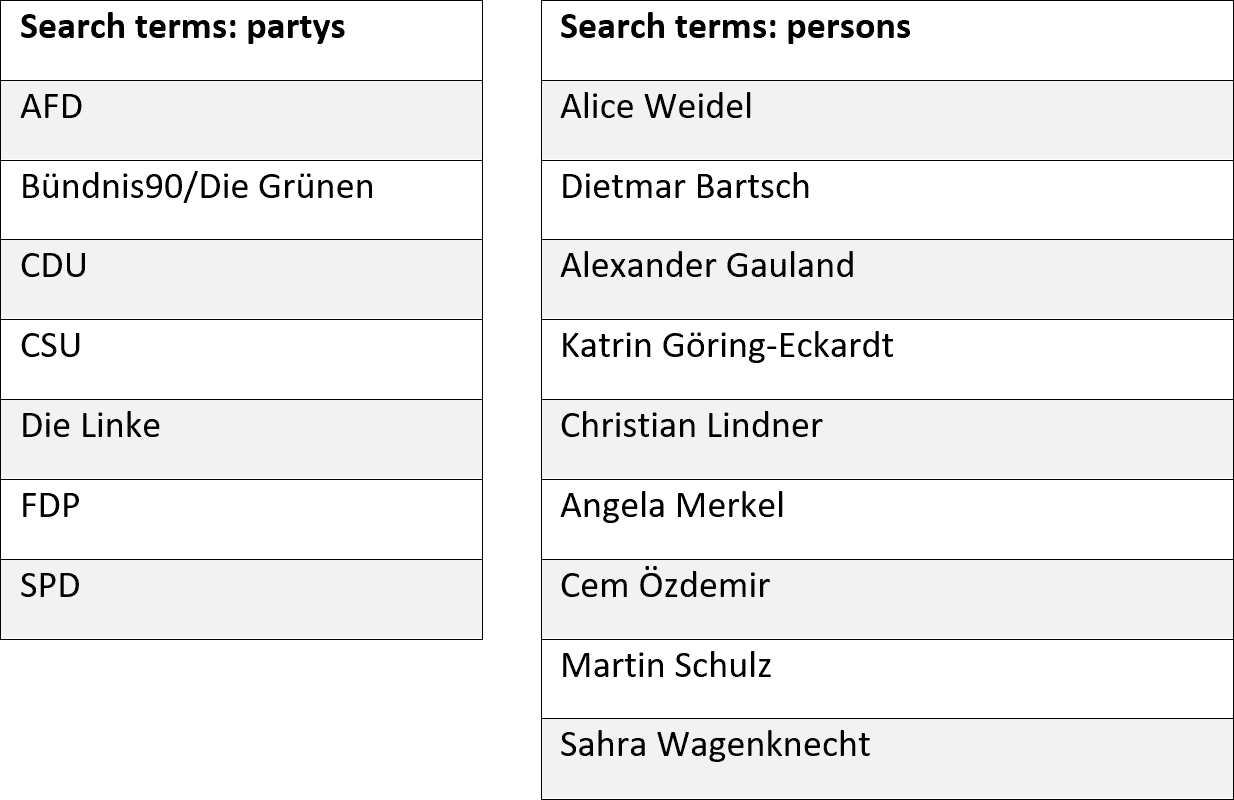}
    \caption{The plug-in's search terms for fixed search times.}
\end{table}

It should also be noted that an automated search for about an approximate dozen of search terms can have an influence on the search engine algorithm itself. On Google Trends, over the runtime of our data collection and for the search terms "Dietmar Bartsch", "Katrin G{\"u}ring-Eckardt" and "B{\"u}ndnis90/Die Gr{\"u}nen" it can clearly be seen that the search request volume was hereby increased (see Figure 2).\\ 
Since the search requests were performed automatically and none of the offered links were actively clicked, we suspect the effects to be rather low. However, lacking exact knowledge of the underlying algorithm, this can't be proven and has to remain unevaluated.

\begin{figure}[ht!]
    \centering
    \includegraphics[width=0.5\textwidth]{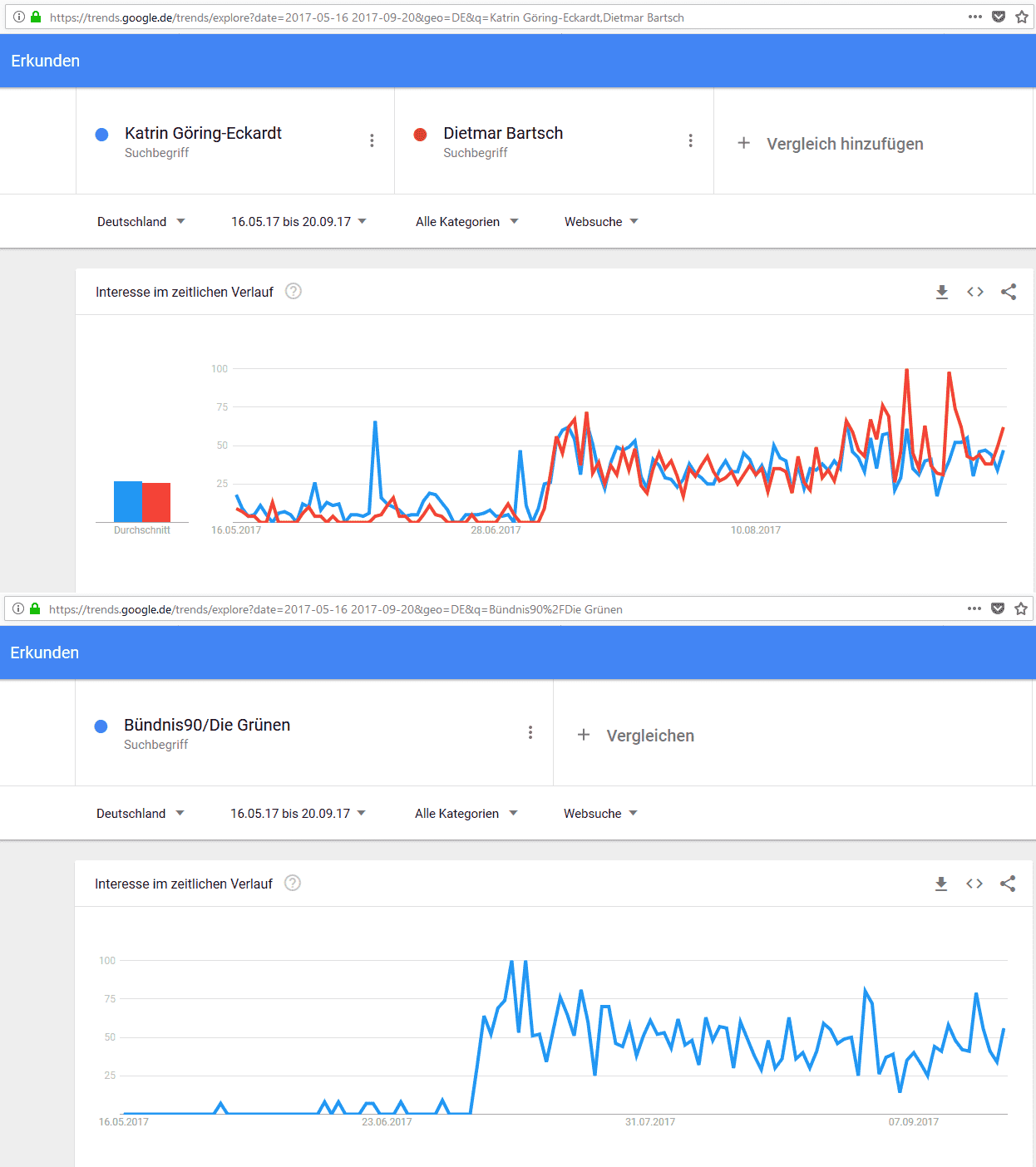}
    \caption{Chronological sequence of the search terms "Katrin G{\"o}ring-Eckardt" and "Dietmar Bartsch" (above) and the search term "B{\"u}ndnis 90/Die Gr{\"u}nen" (below). The Google Trends diagrams clearly show the increase in search occurrences for the search terms due to the plug-in, which was unlocked on July 7th 2017.}
\end{figure}

\subsection{Data structure and relevant terms}
For further analysis, the available data was structured as follows. On the one hand we differentiated between search results directly on Google (www.google.de) or on the search engine provider's news page (news.google.com). The entire database was divided into these two categories. While usually 20 results according to the search terms were displayed on the news portal, 10 results as well as three additional so called top stories are displayed to the user using the default search feature (see Figure 3).\\
While Google searches primarily refer to personal websites, social media accounts and aggregate subject/topic pages for the parties and persons (cf. Section 3.4), the Google news search only shows news from previously registered partners\footnote{See Google's help center, e.g. \href{https://support.google.com/news/publisher-center/answer/6016113?hl=en&ref_topic=9010378}{https://support.google.com/news/publisher-center/answer/6016113?hl=en\&ref\_topic=9010378}}. As a result, the lifespan of news results is limited, i.e. that most news are displayed to users over a low number of search points in time, while the personal websites and social media accounts of parties and politicians are almost always shown. Thus, a separate inspection of news search and Google search results seems reasonable.\\
For regular Google search results we drew a distinction between the (not always received) top stories and "organic" search results, i.e. the 8-10 results in the lower left segment of the results page (Figure 3 shows two organic search results for the search term "Angela Merkel"). At times, Google's search result pages contain information in the right segment as well, e.g. advertisements or info boxes regarding individuals or parties. Those were not transferred to our servers, but rather potential top stories and the actual search result page.\\
Apart from the search type ("Google", "Google news search") the following information was saved:
\begin{itemize}
\item An approximate location based on the user's transmitted IP address.
\item The user login status in their Google account. A user can be "logged in" or "not logged in".
\item The browser language (not the search language that is entered in the Google account).
\item	The search term and time stamp of the search.
\item	An ID generated by the plug-in that does not give any hints regarding the user, but remains unchanged for all data donations, as long as the plug-in is not re-installed.
\item	If available, a descriptive link text is saved as well (usually only available for organic search results, but not always).
\item	If it is a top story, the corresponding date (e.g. "54 minutes ago", "3 hours ago") is saved as well.
\item	The search results' URL.
\item	Most of the time, for top stories and Google news results the medium (news source) and the news title are stated and saved (e.g. "Dresdner Neueste Nachrichten" with "This is what our readers have to say regarding Cem {\"O}zdemir's appearance").
\end{itemize}
Additionally, the following terms are used in this report:
\begin{itemize}
\item[] {\textbf{Investigation period}: This term is used for the time period between August 21st 2017 and September 24th 2017. Here, only weekdays and the election weekend specifically are taken into account (for further information, see Section 3.1). Therefore, the investigation period spans 27 days.} 
\item[] {\textbf{Search time}: We define the search time as a day and corresponding time of day within the investigation period, which can be 12:00, 16:00 or 20:00. We perform this limitation because the number of searching users is significantly lower for other times of search. The sum of times of search thus amounts to 81 (three separate times for each of 27 days in total).}
\item[] {\textbf{(Search) result list}: We understand a result list as the sum of URLs which are delivered to a user for a given search term and search time.}
\item[] {\textbf{Top stories and organic search results}: The top stories term is used for up to three news items which Google occasionally displays at the top of a regular Google search request (see Section 6.1). Apart from purely textual information, a corresponding image is shown (see Figure 3). The remaining search results will hereinafter be referred to as organic search results.}
\item[] {\textbf{Top-level domain}: Each URL references a main domain (top-level domain) and directories potentially located there. The main domain corresponds with the portion between the URL protocol specification (http://, https://) and the first following slash ("/"). The top-level domain of http://www.faz.net/\\aktuell/wirtschaft/gruenen-chef-cem-oezdemir-will-\\gelaendewagen-bestrafen-wer-suv-faehrt-soll-die-kosten-fuer-die-umwelt-tragen-15201893.html consequently is www.faz.net.}
\end{itemize}

\begin{figure}[ht!]
    \centering
    \includegraphics[width=0.5\textwidth]{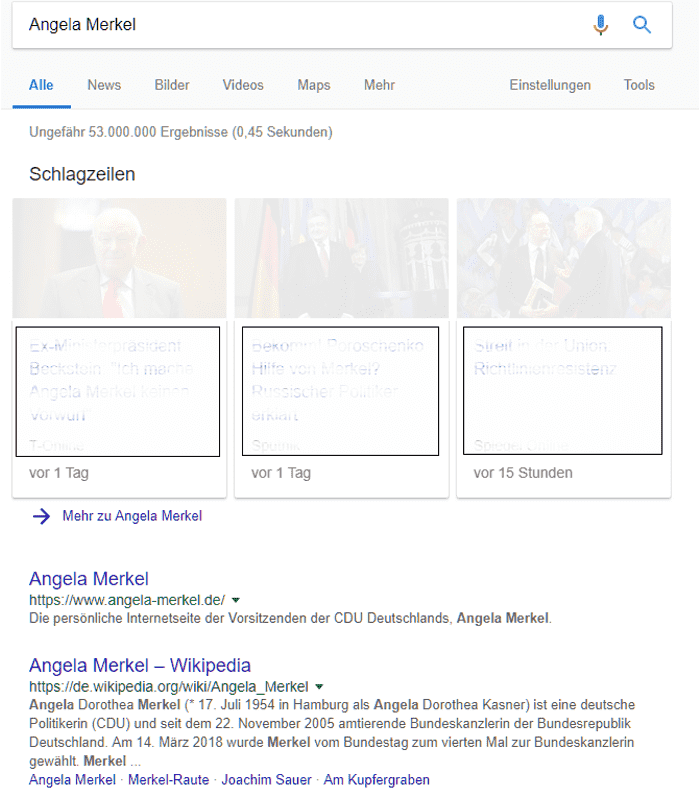}
    \caption{Google search for Angela Merkel with three delivered top stories.}
\end{figure}

As is the case with most data collections, corrupt or divergent entries occur. The plug-in, too, did not run smoothly from the onset and produced partially corrupt data. Therefore, we describe the necessary data preparation below.
\subsection{Data preparation}
The Firefox version's first plug-in assigned the same ID to all users. As we intended to analyze the changes to search result lists over time, we decided to generally disregard that data in order to achieve a consistent database. As a result, 34\% of all donated URLs for the Google search and 41\% for Google News searches are omitted.\\
A first analysis of the available data (see Section 3) shows further irregularities. Noticeably, the database contained search result lists which in length didn't conform to expected standards (ten entries for the common Google search and 20 entries for searches on the news portal). For instance, the database contained some datasets with 200 entries in the search result lists. This can be traced to the possibility of individually adjusting the number of search results displayed on the first page. We have shortened these lists to the standard 10 plus potentially displayed top stories. Other errors are based on imperfect coding of the first Firefox plug-in, resulting in search result lists which contained the same URL throughout the list. Those lists were not included in our analysis.\\
The same applies to URLs which only contained a reference to the corresponding URL at Google (google.de/url) or displayed a URL entry which merely referred to "google" and didn't contain a full link -- those entries point to a private search result. Moreover, it was noticeable that a series of search result lists contained significant numbers of URLs that referred to websites in other languages.\\
To dispose of foreign results, we manually sorted top-level domains of all websites by language. This way, we determined the share of German websites and kept those whose share was above 50\%. As a consequence, our subsequently used database was limited to German result lists. With these adjusted measures, the datasets in our investigation period for the Google search were reduced by 19,6\%\footnote{From 4.416.585 to 3.564.583} and by 16,7\%\footnote{From 6.712.733 to 5.597.480} for Google news data. A short list of the adjusted datasets can be found in Table 2.

\begin{table}[ht!]
    \centering
    \includegraphics[width=0.4\textwidth]{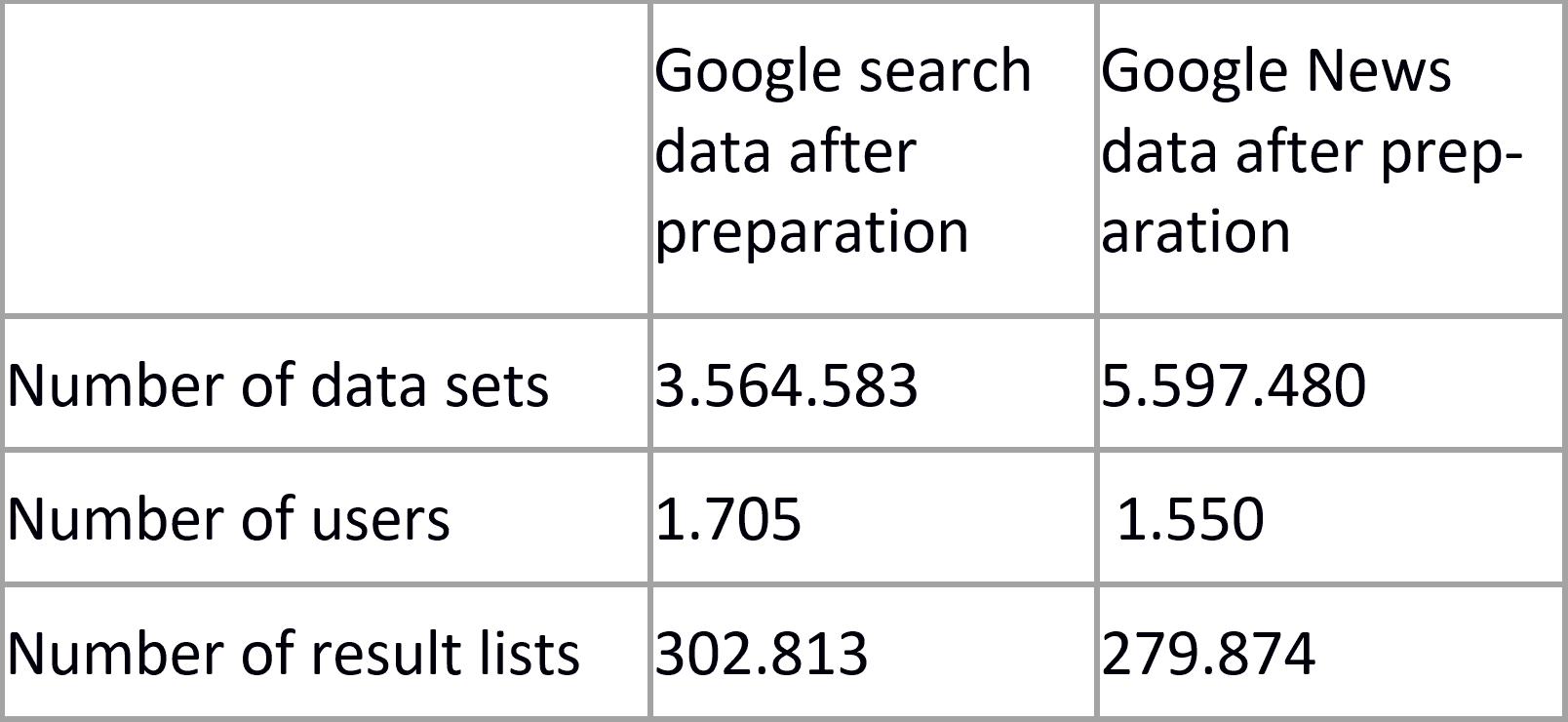}
    \caption{Tabular overview of donated data after data preparation.}
\end{table}

\section{Data Overview}
In this section we give a first overview of the data.
\subsection{Chronological distribution of search requests}
When looking at the daily distribution of our data donations, a clear decrease in donated search result lists during weekends showed. In Figure 4 the number of received URLs per day and search term are displayed and a wave-like pattern can be seen, while the lowest points are located at the weekends. Consequently, significantly fewer users had their browsers open on weekends and donated data, which can be attributed to a lower computer usage on weekends.\\

\begin{figure}[ht!]
    \centering
    \includegraphics[width=0.5\textwidth]{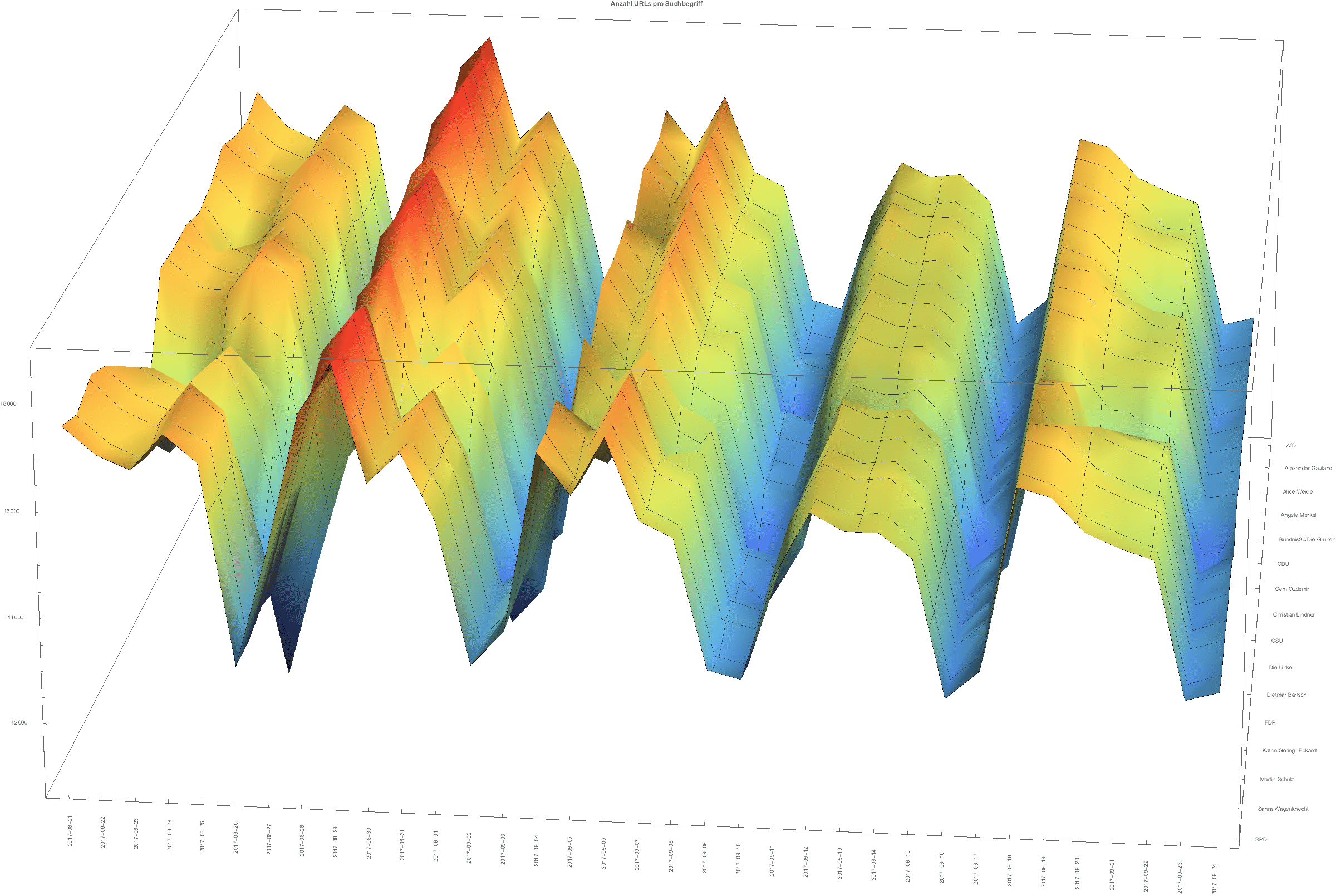}
    \caption{Number of submitted URLs for Google Search during the last 5 weeks before the 2017 federal election.}
\end{figure}

The daily number of data donors were mostly kept stable for both Google Search (see Figure 5) and Google News (see Figure 6) by limiting our observations to weekdays over the investigation period. Thus, between 450 and 550 users were online and donated their search results on the respective days. Also, the ratio between logged in and not logged in users remained similar over the investigation period. The weekend from September 23rd/24th was only added due to its proximity to the election and appears slightly out of the ordinary due to its 350 users. Statistically speaking, the numbers of cases are still sufficiently high to obverse those days as well.\\

\begin{figure}[ht!]
    \centering
    \includegraphics[width=0.5\textwidth]{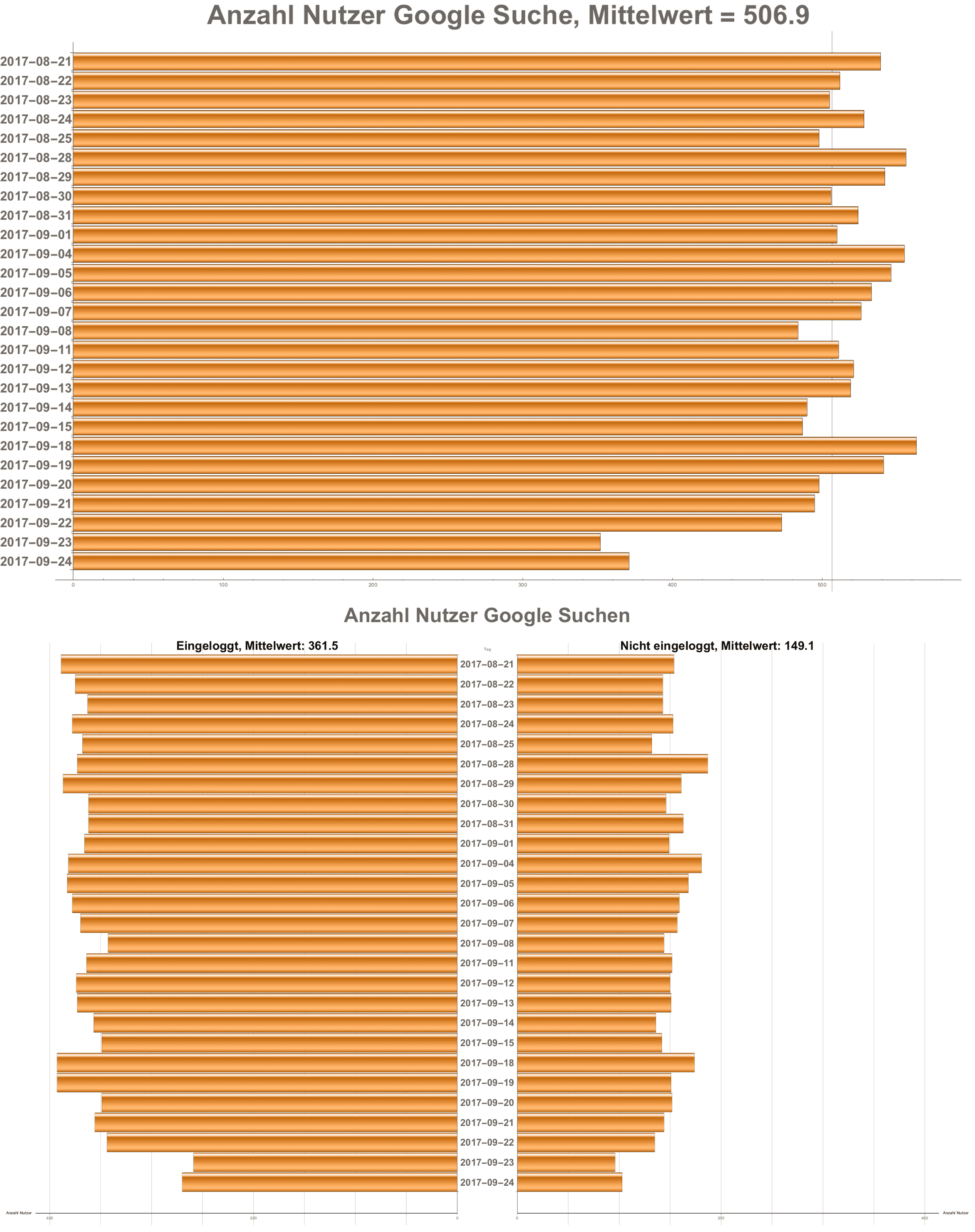}
    \caption{Number of users on the respective days of the investigation period for Google Search, after data cleansing.}
\end{figure}

\begin{figure}[ht!]
    \centering
    \includegraphics[width=0.5\textwidth]{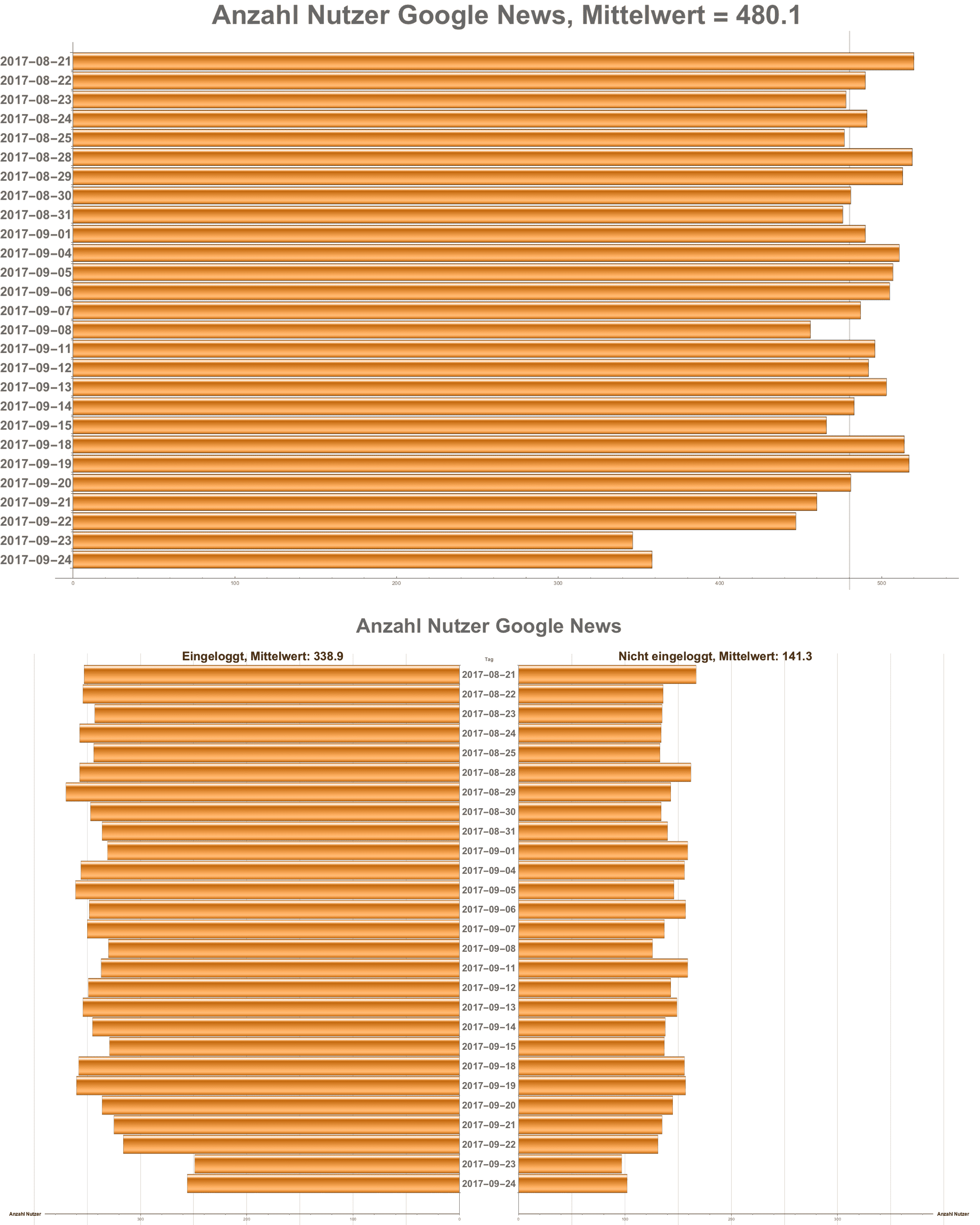}
    \caption{Number of users on the respective days of the investigation period for Google News, after data cleansing.}
\end{figure}

\subsection{Geographic distribution of data donors}
The distribution of users on the map of Germany in Figure 7 shows that we managed to acquire data donors Germany-wide and that our data preparation (see Section 2.3) does not suggest regional effects: Locations that are not included after our data preparation are displayed as reds dots. Those are spread all over Germany; overall, the West of Germany is represented more strongly.

\begin{figure}[ht!]
    \centering
    \includegraphics[width=0.5\textwidth]{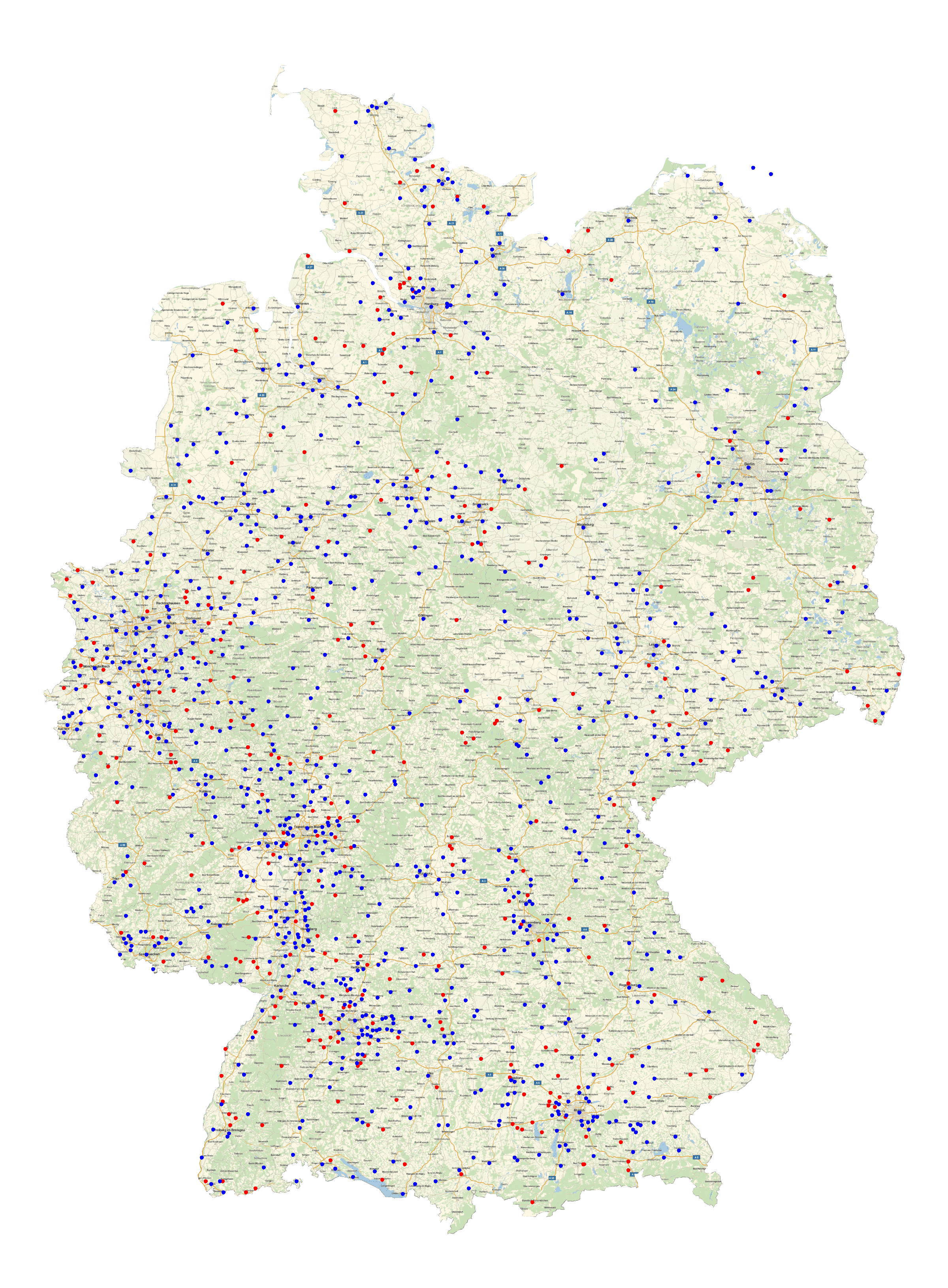}
    \caption{Distribution of data donors in Germany, where the respective positions were determined on the basis of the IP address and are therefore only an approximation of the position of the individual users. The red dots are coordinates that are no longer included in the database after data cleansing.}
\end{figure}

\subsection{Distribution of internet and media offerings in the search results}
For the remaining data set we first calculated which top-level domains managed to get onto the first search result page most often for every search term. Figure 8 shows the results for the political parties and Figure 9 those for the politicians. For all parties except for Die Linke the respective party's own top-level domain ranks first, for Die Linke it ranks 4th, while the parliamentary group's website ranks second. Likewise, the respective Facebook account is always the second to fourth most common link, Twitter manages an entry into the top 10 for four parties; for all parties except for the AfD, this is accompanied by the German Wikipedia entry. For the AfD there is a German Wikipedia entry as well, but it does not appear as often as the aggregate topic pages of Focus, Zeit, Merkur, Spiegel, FAZ and Tagesspiegel. A regular appearance for parties except for Die Gr{\"u}nen is the top-level domain www.bundestagswahl-bw.de, which offers an overview of all parties. "bundestageswahl-2017.com", which appears for Die Gr{\"u}nen and the SPD, is a private collection of information relating to the parties and their platforms. It stands out that both B{\"u}ndnis 90/Die Gr{\"u}nen and Die Linke predominantly have their own websites in the top 10 results, or those that could be altered by them with some degree of effort, e.g. in case of false coverage. This includes Facebook and Twitter as well as Wikipedia in general.\\
For politicians, the personal website is always in the top 10 as well, for Alexander Gauland and Alice Weidel those are subpages of the top-level domain www.afd.de and thus not specifically visible. The corresponding German Wikipedia entry appears for all of them as well, besides a changing number of social media accounts, which are completely absent for Alexander Gauland. As a matter of fact, we could not find any personal social media account of Alexander Gauland, his website refers to the respective accounts of the AfD.\\
Most big German online news magazines have topic pages, which collect news relating to a person or institution and often display an introductory, general text about the topic. Such person-related topic pages can be found for politicians among the top 10 of the top-level domains, but news from those and other sources as well.\\

\begin{figure}[ht!]
    \centering
    \includegraphics[width=0.5\textwidth]{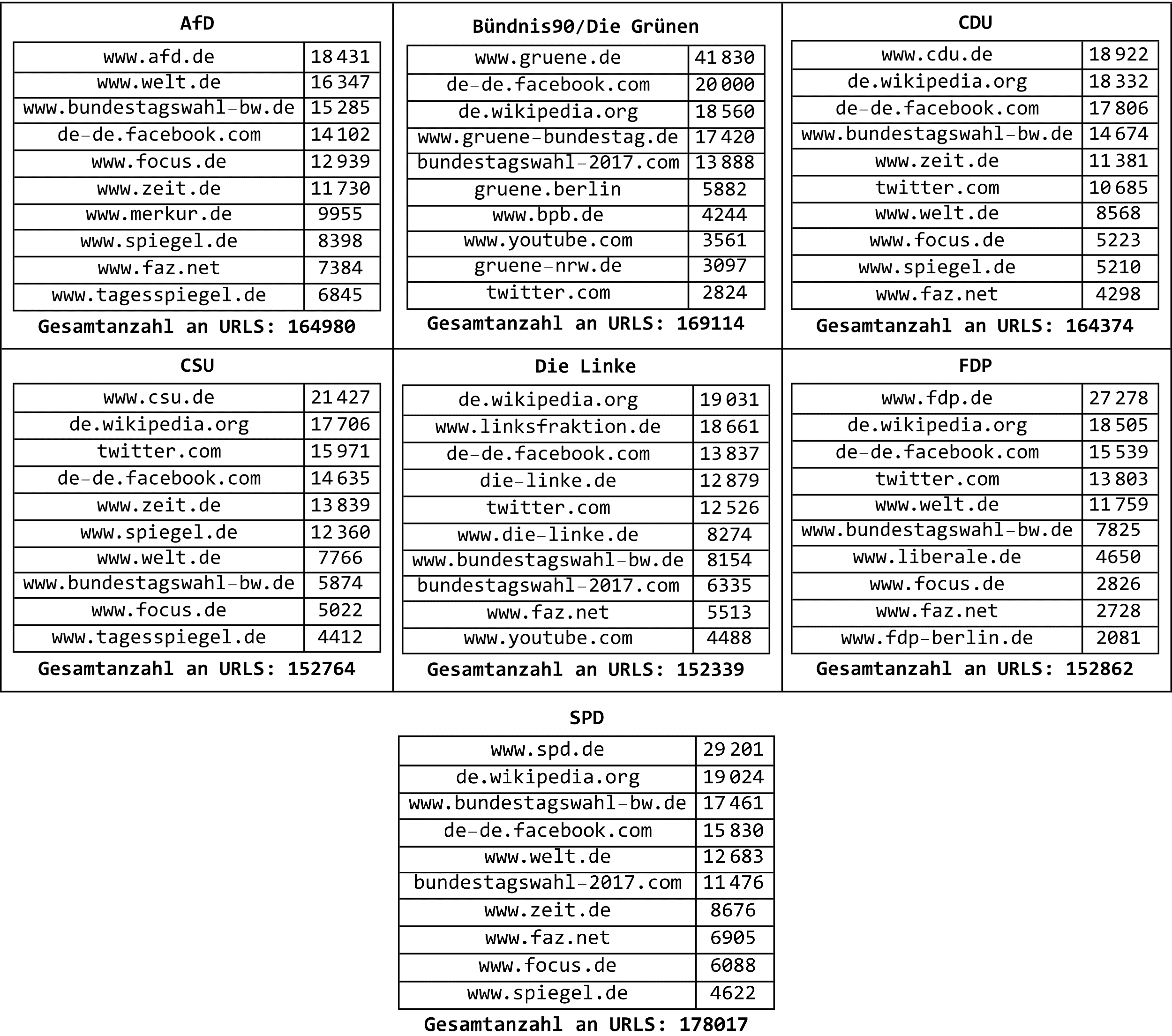}
    \caption{Most common top-level domains for the organic Google search for the parties.}
\end{figure}

Examining the top-level domains for the top stories, a less uniform picture emerges (see Figure 10): While AfD, CDU, CSU, SPD and FDP receive most top stories from far-reaching media companies, the top 10 sources for Die Gr{\"u}nen and Die Linke are lesser-known in some instances. For the searched persons, there are other sources to be found besides the most far-reaching, like the Generalanzeiger Bonn (for Cem {\"O}zdemir) or www.epochtimes.de (for Alexander Gauland and Katrin G{\"u}ring-Eckhardt) (see Figure 11). For the sake of completeness, Figure 12 and 13 also show the top 10 of top-level domains for the respective searches on Google News. A media studies classification of the respective sources can not be provided here -- the data however is available for future analysis\footnote{https://datenspende.algorithmwatch.org/data.html}.\\

\begin{figure}[ht!]
    \centering
    \includegraphics[width=0.5\textwidth]{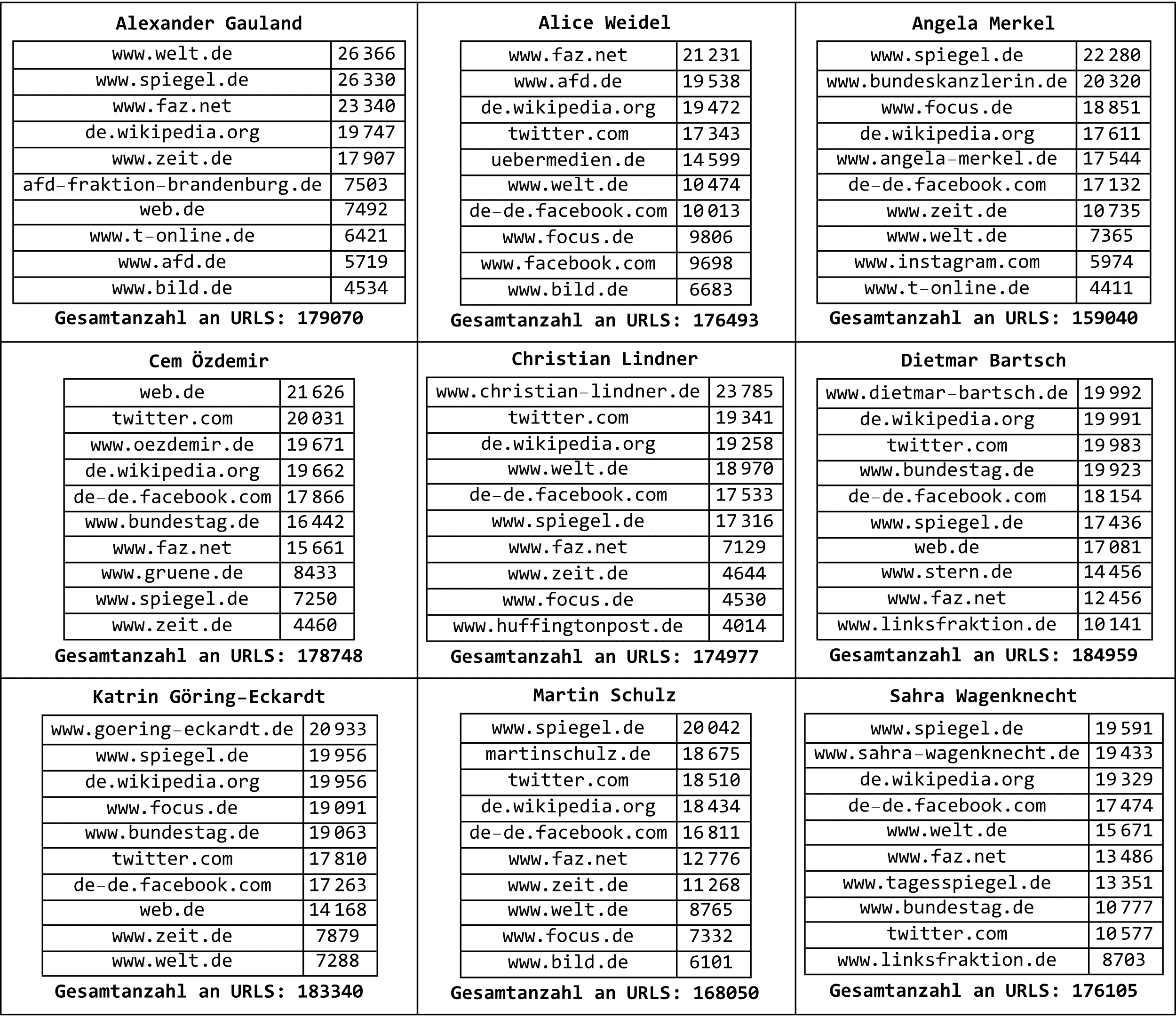}
    \caption{Most common top-level domains for the organic Google search for the persons.}
\end{figure}

Overall, the first result page both for Google News and Google's organic search are dominated by renowned media companies, especially from the printing sector. Exceptions to this are the online newspaper Huffington Post and Freemail service provider t-online.de, which often appear as news sources. When searching for the terms "Christian Lindner" and "FDP", Google provides a website as a news result which is being managed by the party's federal branch.

\begin{figure}[ht!]
    \centering
    \includegraphics[width=0.5\textwidth]{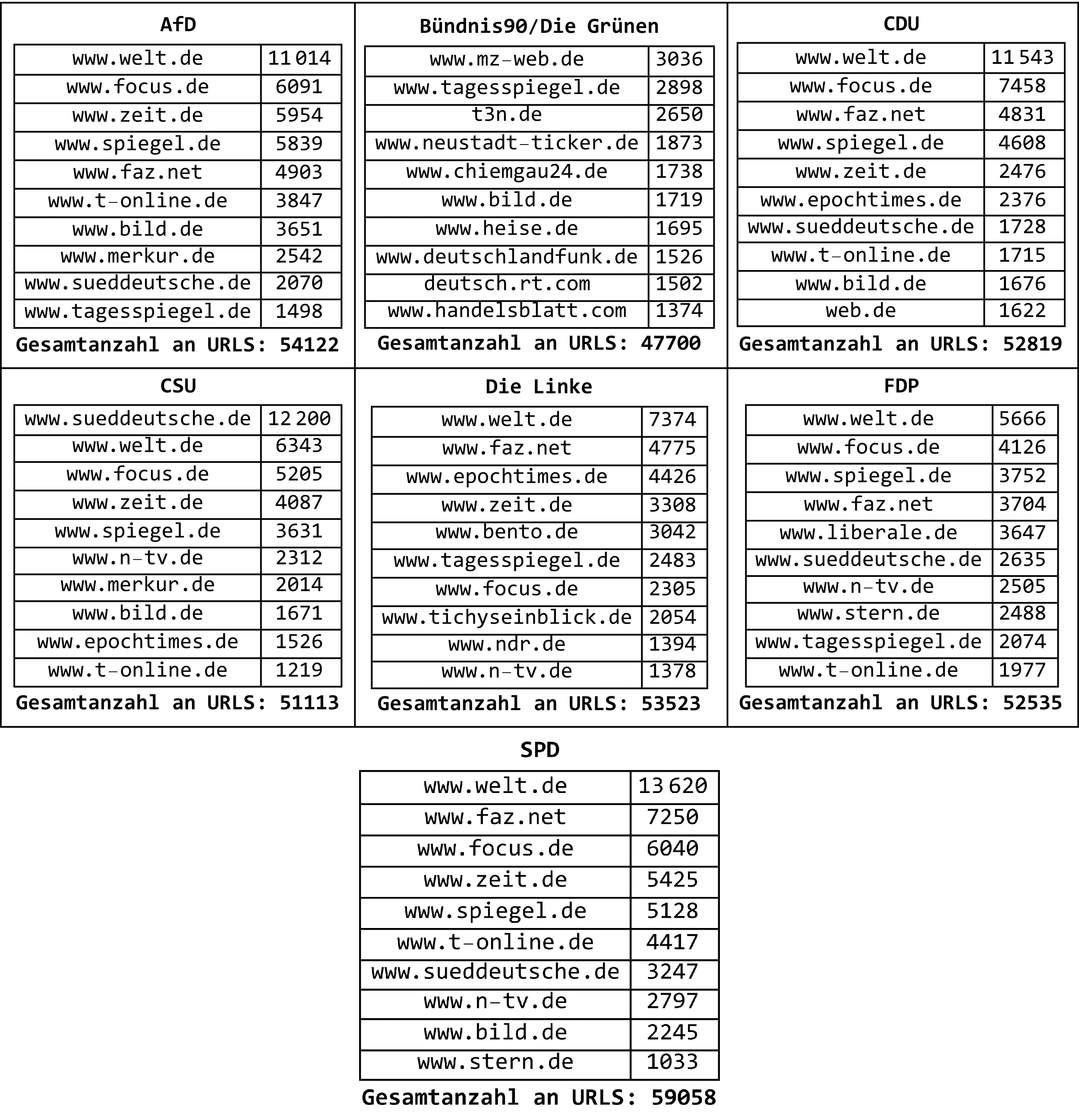}
    \caption{Most common top-level domains for the top stories of the Google search for the parties.}
\end{figure}

\begin{figure}[ht!]
    \centering
    \includegraphics[width=0.5\textwidth]{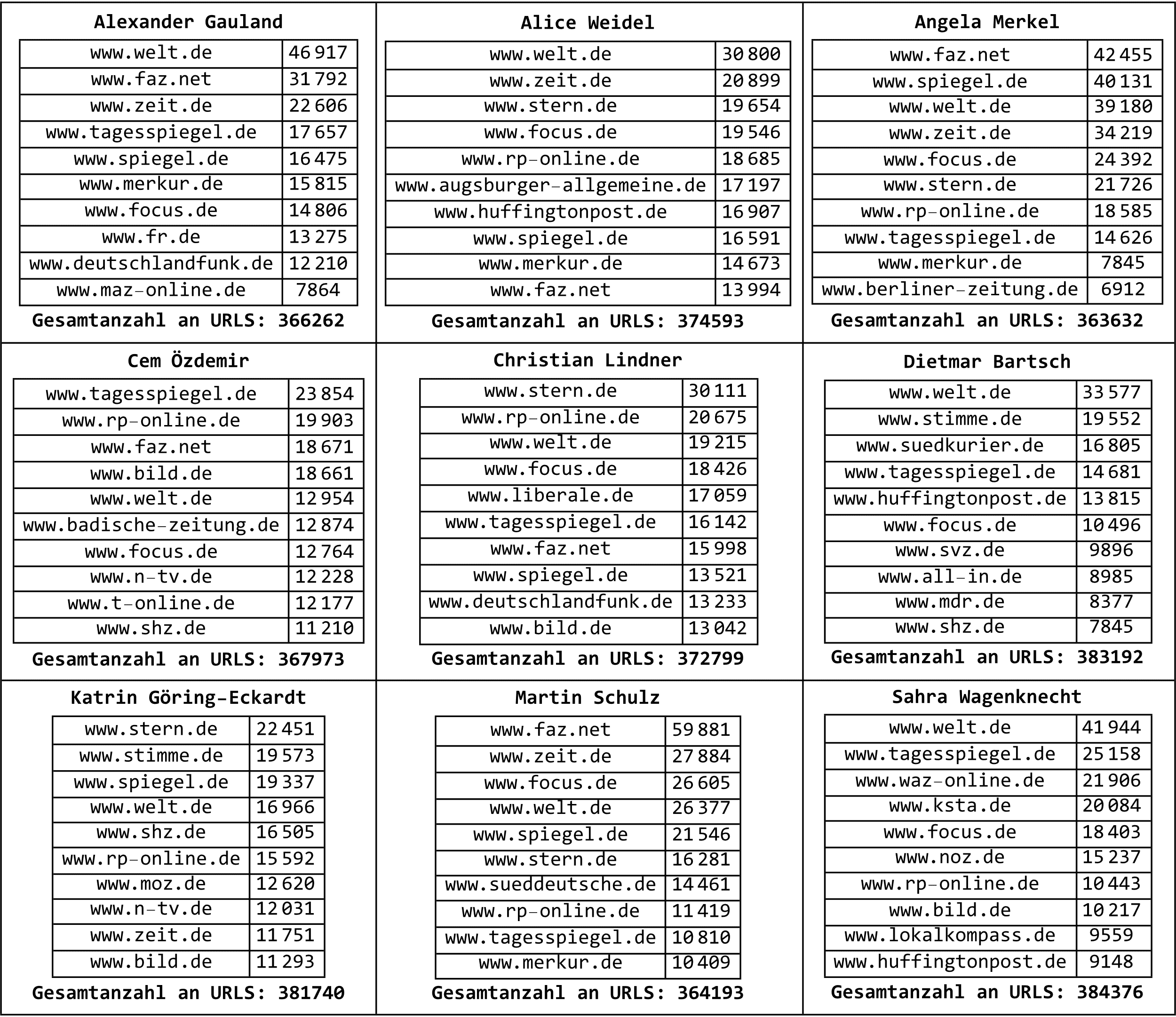}
    \caption{Most common top-level domains for the Google News person searches.}
\end{figure}

\begin{figure}[ht!]
    \centering
    \includegraphics[width=0.5\textwidth]{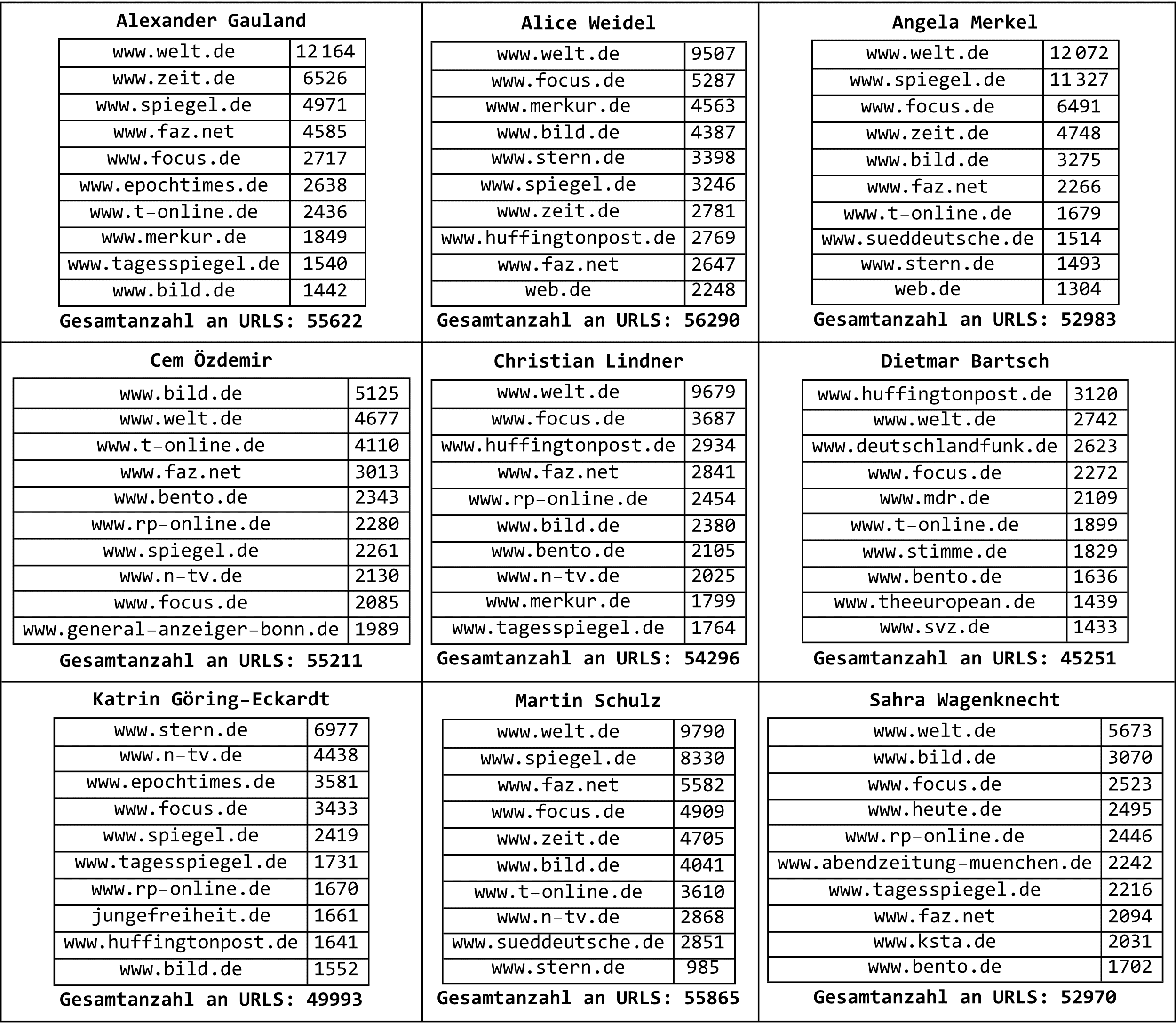}
    \caption{Most common top-level domains for the top stories of the Google search for the persons.}
\end{figure}

\begin{figure}[ht!]
    \centering
    \includegraphics[width=0.5\textwidth]{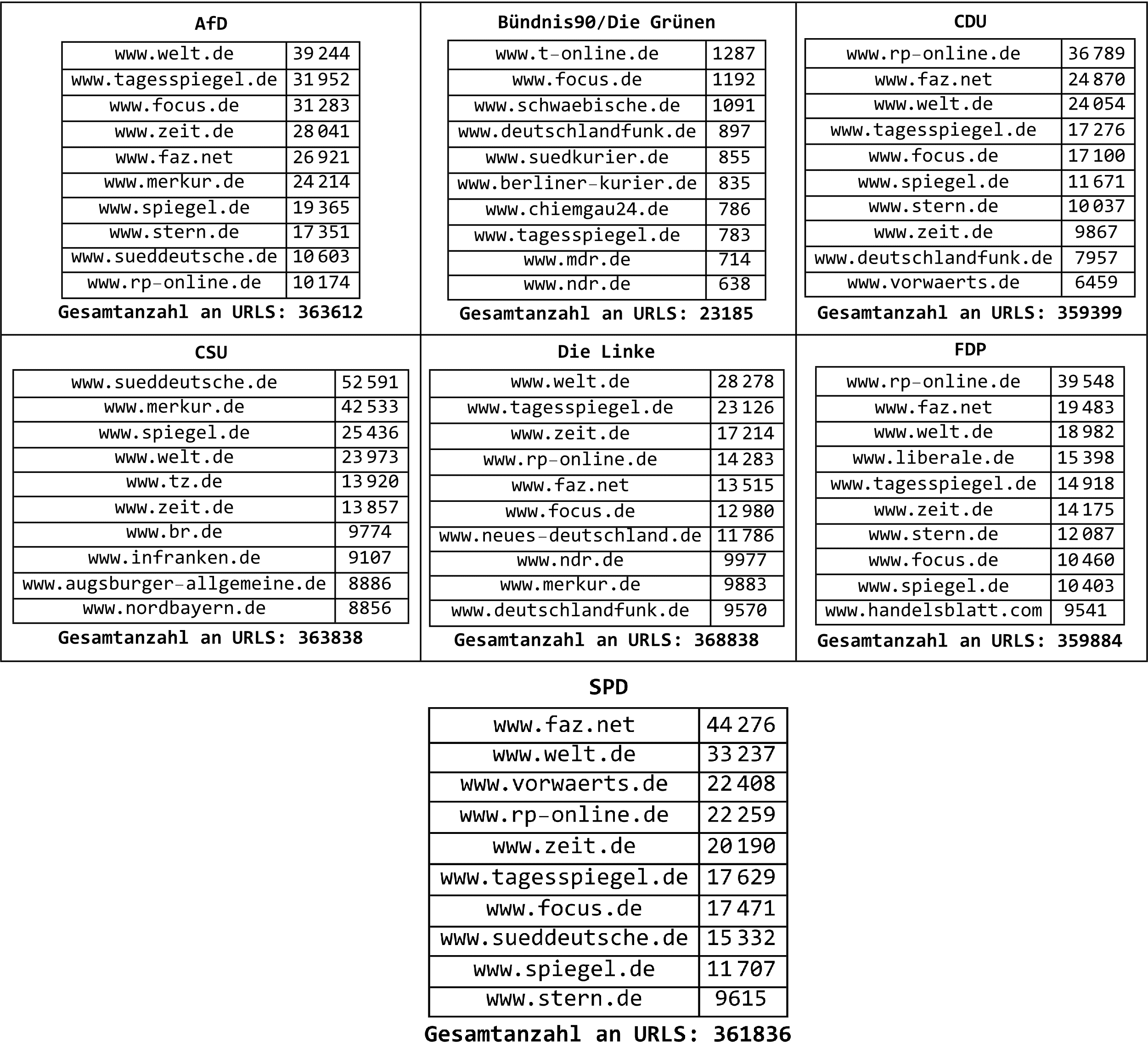}
    \caption{Most common top-level domains for the Google News party searches.}
\end{figure}

\subsection{Owned Content, Social Media and media offerings}

\begin{figure}[ht!]
    \centering
    \includegraphics[width=0.5\textwidth]{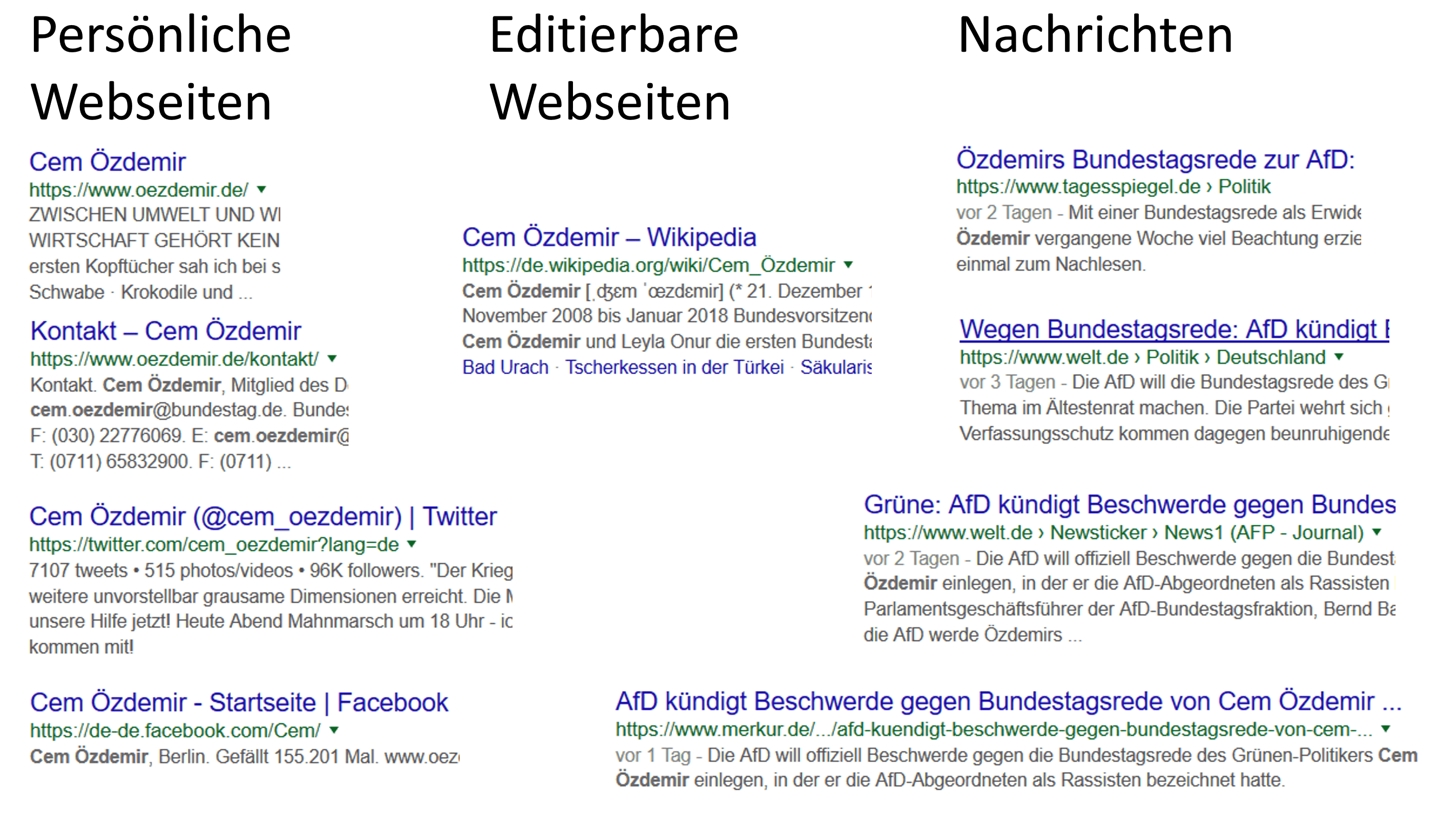}
    \caption{Division of organic search results for the search term "Cem {\"O}zdemir" into three possible categories: personal websites, which can be further divided into "Owned Content" and "Social Media", editable websites such as Wikipedia and news. Besides, aggregate news services such as Freemail portals appear frequently (t-online, web.de), but were not among the results of this search.}
\end{figure}

The search result lists contain different result categories, as Figure 14 shows for an example of a search for "Cem {\"O}zdemir".
An interesting question when assessing the diversity of results delivered by search engines is to look at their distribution on different offering categories and especially in asking the question of the extent to which persons or parties are able to edit their contents. Websites of candidates and parties evidently belong to the latter. To that end, each individual URL was manually assigned to the database and one of 7 different categories, according to its top-level domain; URLs of the Media category were categorized in a more nuanced way\footnote{We thank the Bavarian Regulatory Authority for Commercial Broadcasting (Bayerische Landeszentrale f{\"u}r Neue Medien, BLM) for this difficult work.}. A detailed description of those individual categories can be found in Appendix A.\\

\begin{table}[ht!]
    \centering
    \includegraphics[width=0.3\textwidth]{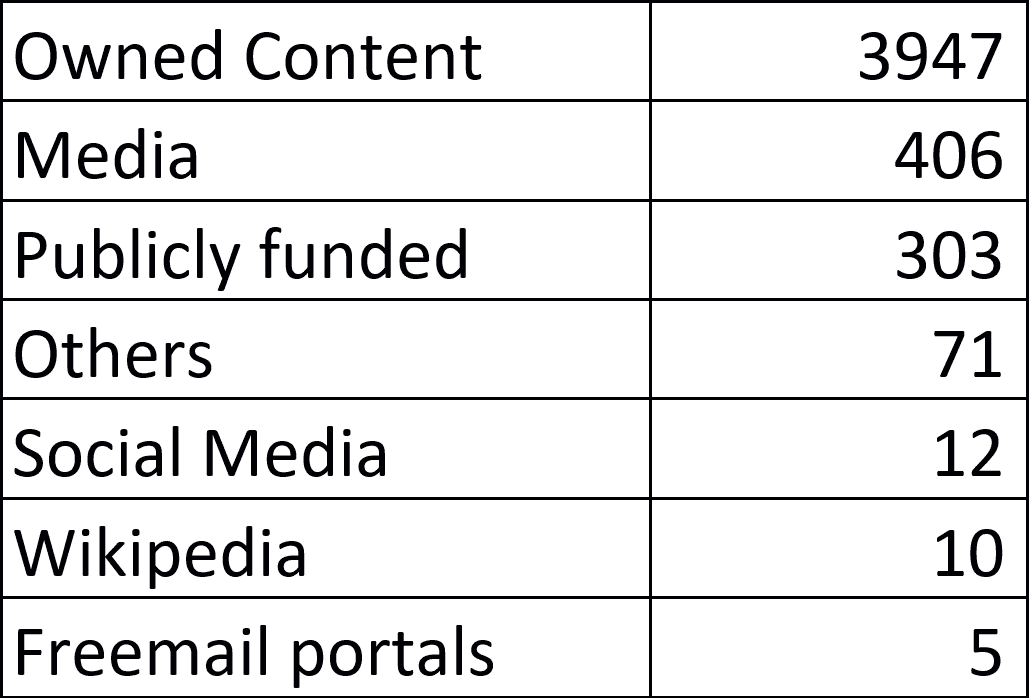}
    \caption{Distribution of all top-level domains of the Google search over the offering categories.}
\end{table}

It is more than astonishing that of the 4754 top-level domains of the Google search that were delivered during the investigation period a predominant portion of 83\% are subject to editability by parties or persons ("Owned Content"). The vast majority of Owned Content consists of pages of local branches belonging to the respective parties (see Table 4). Some parties are significantly more active than others: While the sum of search results for the AfD provides a mere 6 clearly discernible individual top-level domains, it is more than 1000 for the SPD and nearly 950 for the CDU. Besides websites of parties and persons and the approximately 400 media domains, the relatively high number of publicly funded domains stands out (about 300 domains). Those primarily represent cities and municipalities; however, the website of the Federal Agency for Civic Education and the Bundestag are included here as well.\\

\begin{table}[ht!]
    \centering
    \includegraphics[width=0.5\textwidth]{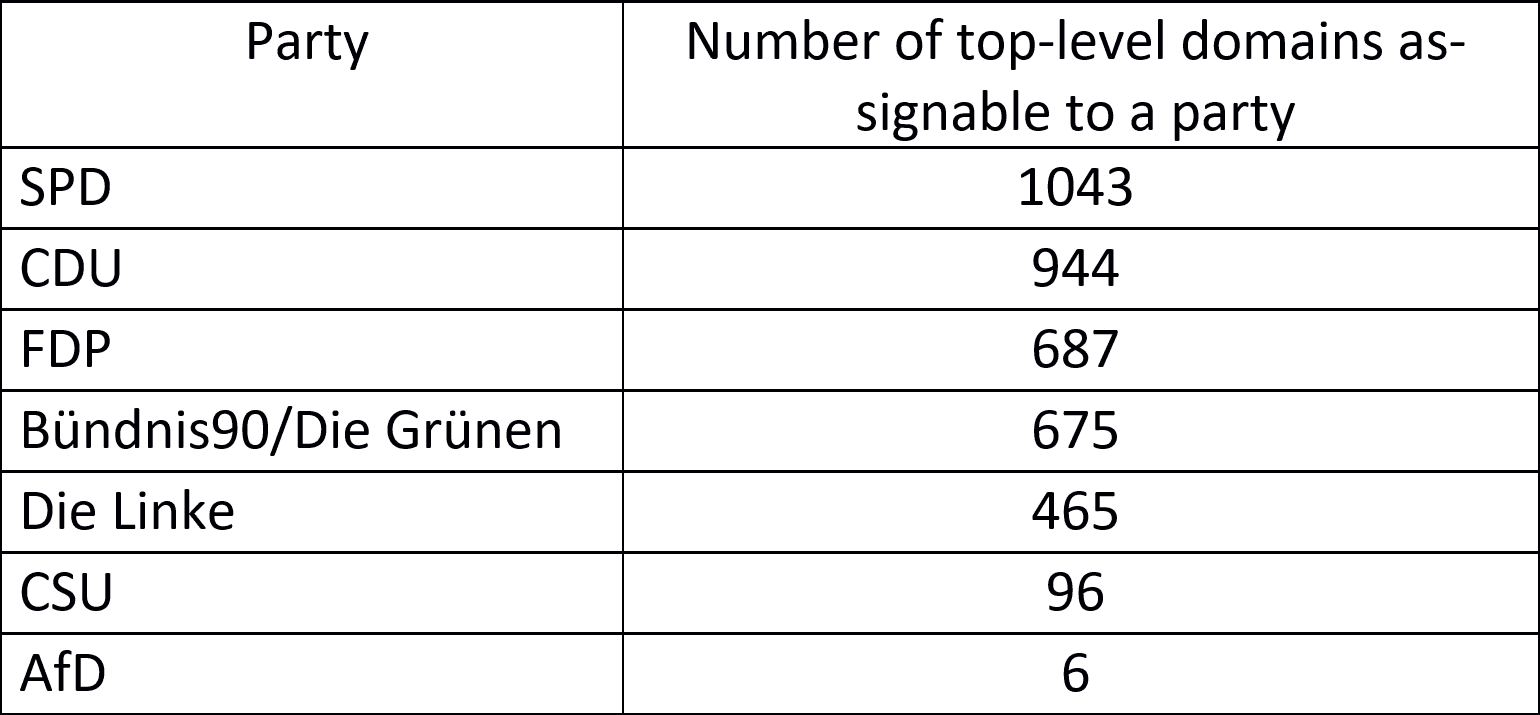}
    \caption{Number of top-level domains which can be assigned to a party. They usually represent a local branch, the party itself or individual party members.}
\end{table}

Figure 15 shows the distribution of organic search results (without top stories) over the entire investigation period. Here, distinct differences can be found regarding the share of URLs in the categories Owned Content and Media between individuals and parties. The amount of links referencing a Wikipedia page lies above 10\%, which means slightly above one link, as there are users who received both the German and English Wikipedia entry. Due to this categorization it is now possible to differentiate between the entire proportion editable by a party or individual and the proportion that can not be edited (media, Freemail portals, publicly funded and others).\\

\begin{figure*}[ht!]
    \centering
    \includegraphics[width=1\textwidth]{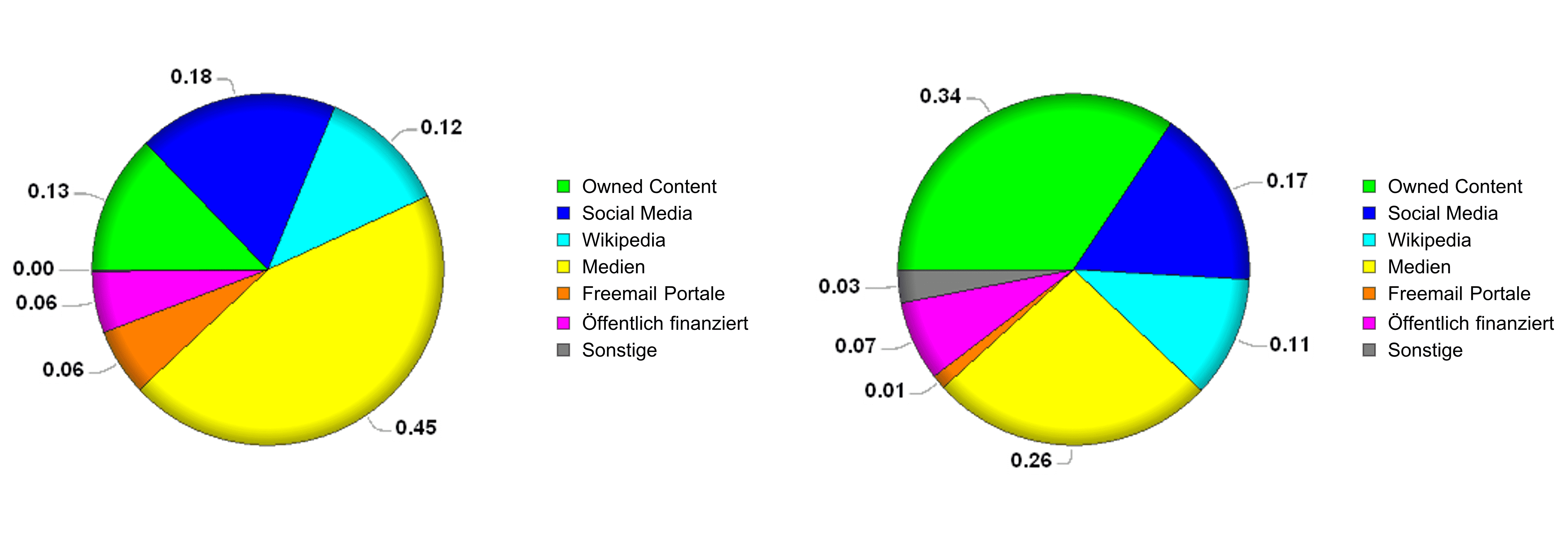}
    \caption{Distribution of organic search results among different categories in \%, left for individuals, right for parties. The 9-10 organic search results a user receives when searching for a person thus contained e. g. on average two URLs from the category "Social Media" and 4-5 links from the category "media". }
\end{figure*}

Summarizing the proportion of the respective categories for the particular parties (see Table 5), further evidence of a varying media presence in the digital space presents itself. While B{\"u}ndnis 90/Die Gr{\"u}nen with 60\% and Die Linke as well as the FDP with about 50\% have achieved a very high proportion of Owned Content (i. e. links to own domains of parties, local branches or politicians), the CDU and SPD manage a mere 30\%. A chicken-and-egg problem arises: A well-working search engine optimization (SEO) provides good visibility of contents for all search engines, but good visibility in search engines can also lead to high popularity of contents in all search engines, which in turn results in good visibility in search engines. Therefore, the reason for some parties' own content being more visible to themselves does not seem assessable. After all the conducted investigations we consider it most likely that those parties are most visible that provide a sufficient amount of timely content both on the regionalization as well as the news level. Table 5 shows the same analysis for search results that represent individuals. Here it is noticeable that especially for Gauland, who does not own social media accounts, less directly or in principle editable domains are delivered in total -- for most other persons approximately two social media accounts are delivered. Frequently a publicly funded domain is shown for those with a seat in the German Parliament (e. g. www.bundestag.de, www.bundeskanzlerin.de), which are missing for Gauland, Weidel, Schulz and Lindner.\\

\begin{figure}[ht!]
    \centering
    \includegraphics[width=0.5\textwidth]{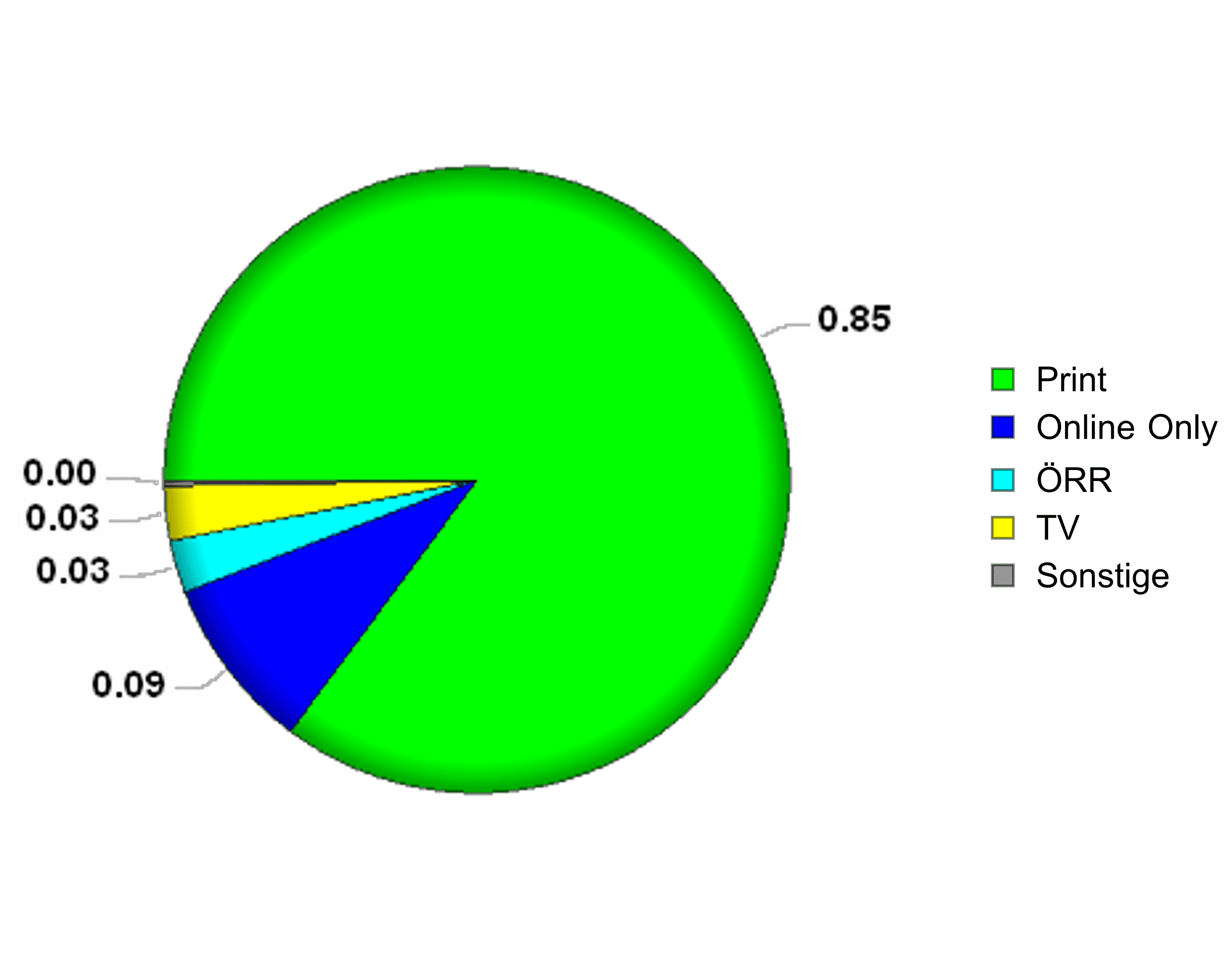}
    \caption{Share of media delivered by Google in the top stories during the investigation period for Google search of all search terms.}
\end{figure}

Finally Figure 16 shows the distribution of categories for the delivered top stories in the Google search. No significant differences both on the search term level as well as aggregated by parties or persons could be observed. The majority of 85\% always comes from the print media sector with an online presence. Media offerings with the internet as their sole path of distribution are displayed in the top stories with 9\%, while only 3\% represent public service channels and online branches of classic German private television service providers.

\begin{table*}[ht!]
    \centering
    \includegraphics[width=0.9\textwidth]{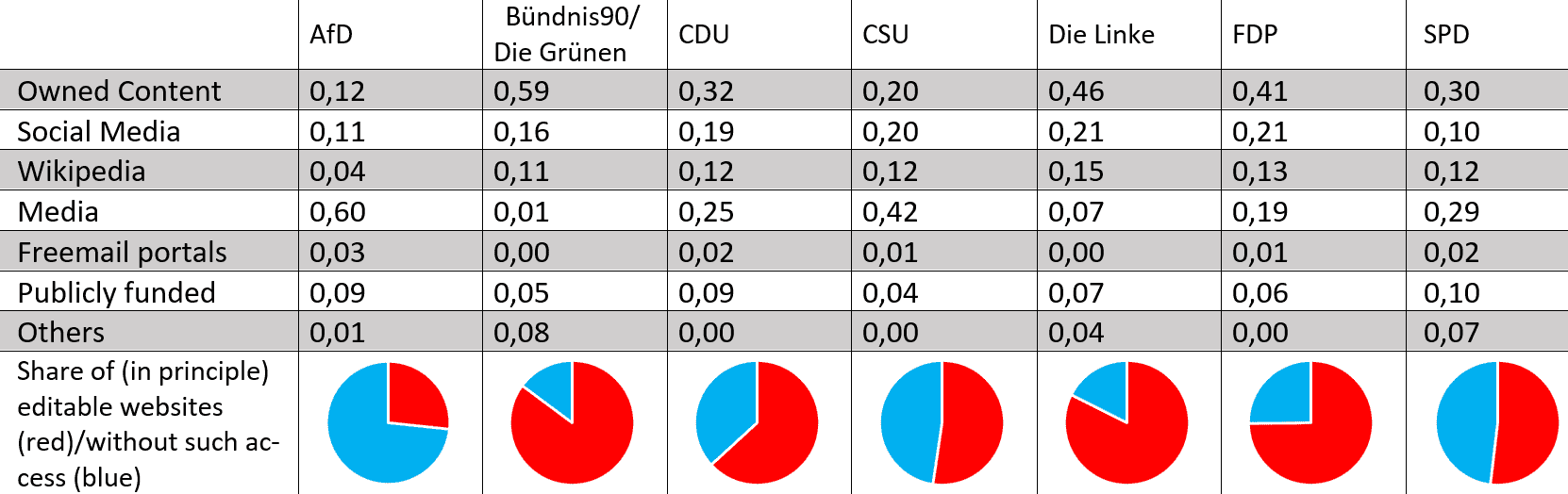}
    \caption{Distribution of the organic search results on the different categories, for the searches for the parties. The circular diagrams represent the proportion of all websites that are (in principle) editable by the person or party in red (Owned Content, Social Media and Wikipedia) and the proportion of websites on which such an intervention is not possible (Freemail portals, media, publicly funded and others), in blue.}
\end{table*}

\begin{table}[ht!]
    \centering
    \includegraphics[width=0.5\textwidth]{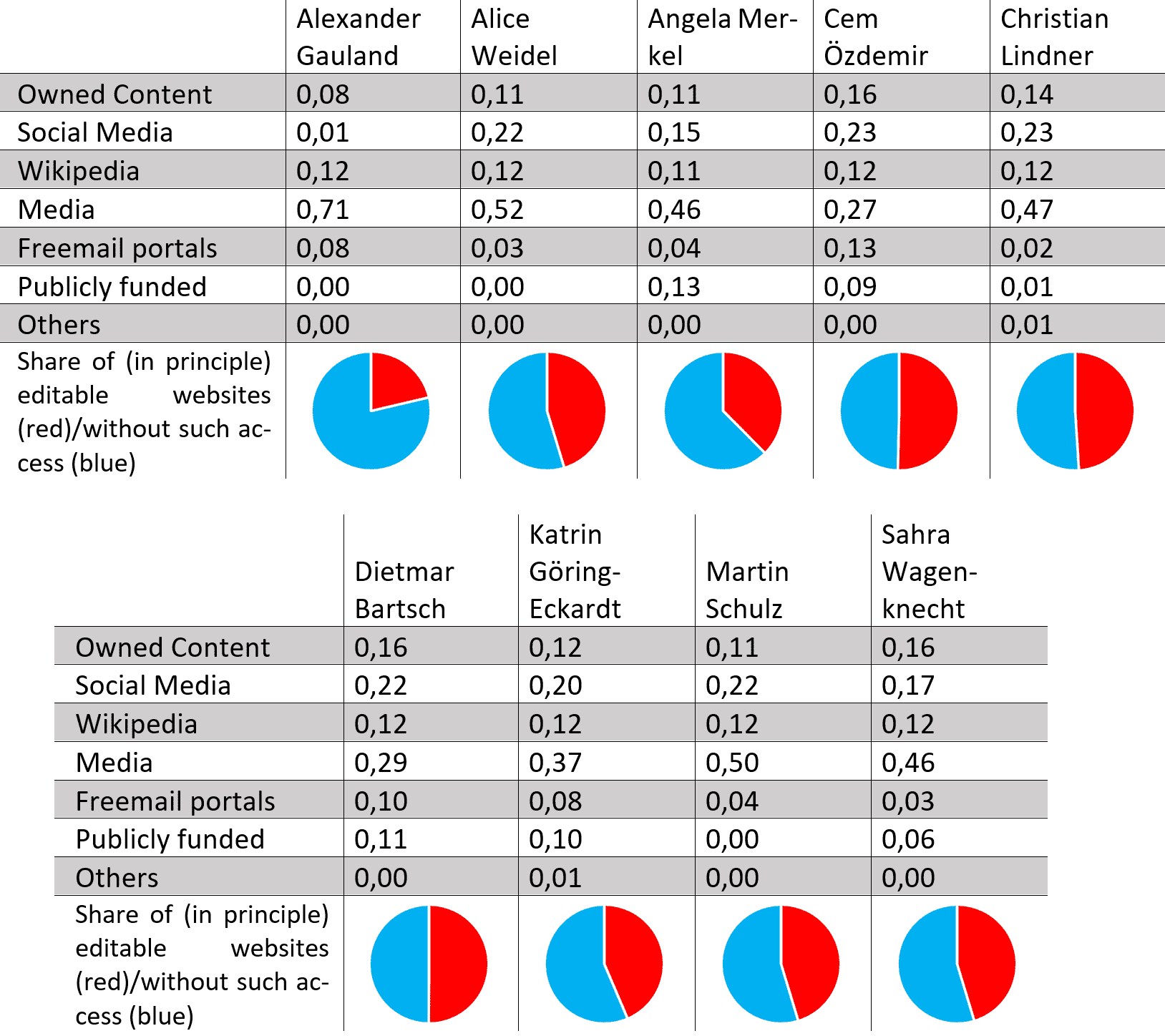}
    \caption{Distribution of the organic search results on the different categories, for the searches for the parties. The circular diagrams represent the proportion of all websites that are (in principle) editable by the person or party in red (Owned Content, social media and Wikipedia) and the proportion of websites on which such an intervention is not possible (Freemail portals , media, publicly funded and others), in blue.}
\end{table}

\section{Scope for personalization/ regionalization}
The \#Datenspende project is a proof-of-concept to demonstrate that black-box analysis enables society to investigate algorithmic decision-making systems such as search engines -- but in principle others such as product recommendation systems or social media as well -- for dangerous spots. A much-debated dangerous spot in recent years has been the algorithm-based formation and reinforcement of filter bubbles based on personalized search and recommendation filters. The question of whether and to what extent search engine users receive different news and information about important political institutions and persons was therefore the focus of the analysis. The fact that we inspected a search engine and not the much more vulnerable social media such as Facebook was above all a matter of easier access to the results.\\
Personalization of search results, as stated previously, is understood to describe a customized preparation or selection of search results that the user has not yet clicked. Suggestions are made based on profiles and click behavior of other users perceived as being similar.\\
The personalization itself can be conveyed in the chosen selection, but also in a different order in the presentation of the found results. In the explanations for the use of the Google service it says:

\begin{quotation}
"We use automated systems that analyze your content to provide you with things like customized search results, personalized ads, or other features tailored to how you use our services. And we analyze your content to help us detect abuse such as spam, malware and illegal content."\footnote{https://policies.google.com/privacy?gl=de\#infouse, (2018, August 12)}
\end{quotation} 

So how significant was this personalization for the data donors before the 2017 federal election?
\subsection{Proportion of similar search results}
The book by Eli Pariser noted the many potential dangers of filter bubbles and expressed the concern that every person is trapped inside their own filter bubble ("being isolated in a web of one" \footnote{From Eli Pariser's TED Talk: https://www.ted.com/talks/eli\_pariser\_beware\_online\_filter\_bubbles, minute 8.}), which likely contains "information desserts" rather than nutritious information \footnote{From Eli Pariser's TED Talk: https://www.ted.com/talks/eli\_pariser\_beware\_online\_filter\_bubbles, minute 5-6.}("information vegetables"). Therefore, a first interesting question is which share of data donor pairs will be shown exactly the same links and in addition, what percentage of them will get these links in exactly the same order.\\
For the results of the Google search (without top stories), the results summarized in Table 7 show the percentage of identical result lists and identical result lists in the same order. In fact, these percentages are essentially in the low single-digit percentage range -- apart from the AfD, where 16\% of data donor pairs contain the same results. It is important to emphasize that this does not mean that 16\% of the result lists are exactly identical. An example: if out of 100 people two groups of people with 5 and 6 members each get exactly the same results and everyone else does not share the exact same links with anyone, then there are 10 pairs of users and 15 pairs of users, each seeing the same links, from a total of 4,950 pairs. This corresponds to 0.51\% of all pairs and thus almost to the result for the CDU.\\

\begin{table}[ht!]
    \centering
    \includegraphics[width=0.45\textwidth]{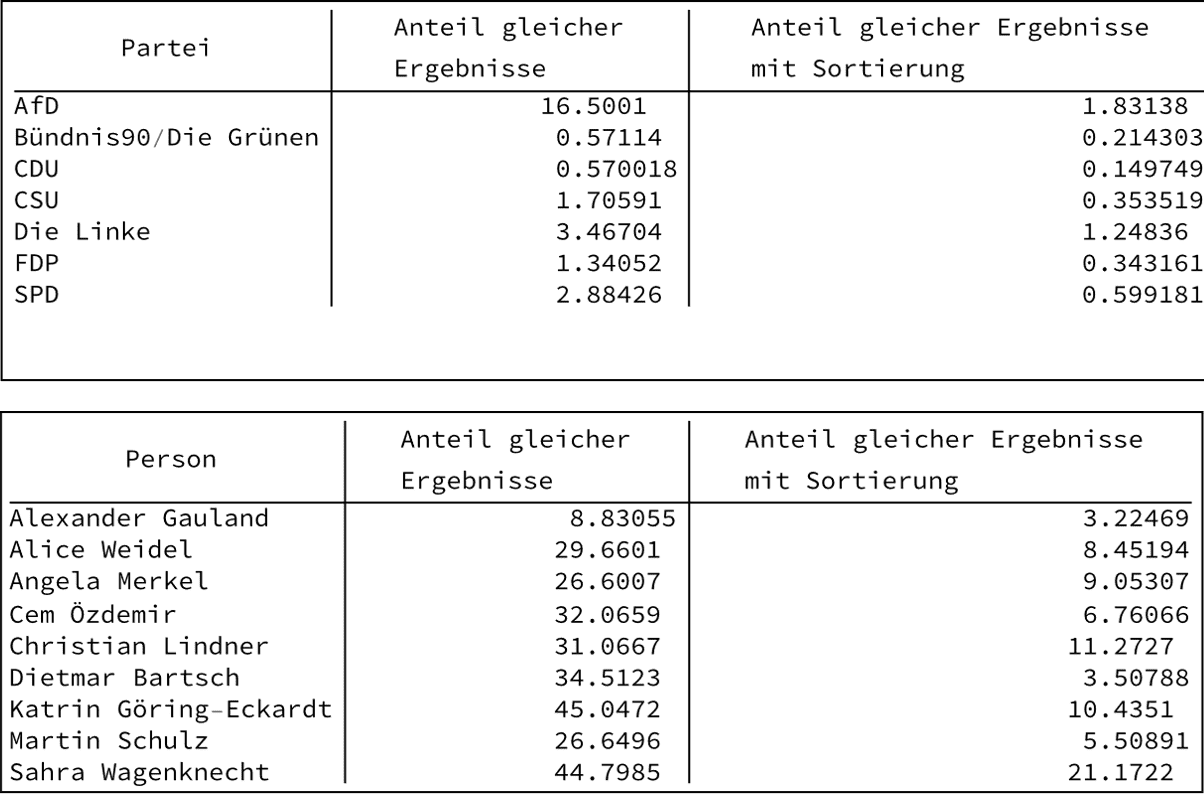}
    \caption{Share (in percent) of identical search results without consideration of sorting, separated by keywords for parties and persons (2nd column). In column 3, the proportion (in percent) of identical search results can be found in the exact same order.}
\end{table}

Truly remarkable is the enormously high share of pairs with similar search results for the persons, which is -- except for Alexander Gauland -- on average at least a quarter and for some almost 50\%. In other words, had we asked any two data donors to do a search for one of the persons at the same time, the same links would have been delivered to a quarter to almost half of those pairs -- and for about 5-10\% in the same order as well. If there were filter bubbles here, at least some of them would be quite spacious and would affect larger groups of people at the same time.\\
A manual inspection of the result lists of Alexander Gauland suggests that notably the missing social media accounts lead to this result: These are replaced by messages which apparently are delivered in a more diverse way. Reviewing Figure 9 supports this hypothesis. The illustration shows that politicians have different numbers of Owned Content and Editable domains among their top 10 top-level domains. Table 8 compares the percentage of similar search result lists to the number of Owned Content and Social Media top-level domains in the top 10 of search result top-level domains. It becomes apparent that more of these top-level domains among the top 10 correlate with a higher conformity of search result lists.\\

\begin{table}[ht!]
    \centering
    \includegraphics[width=0.5\textwidth]{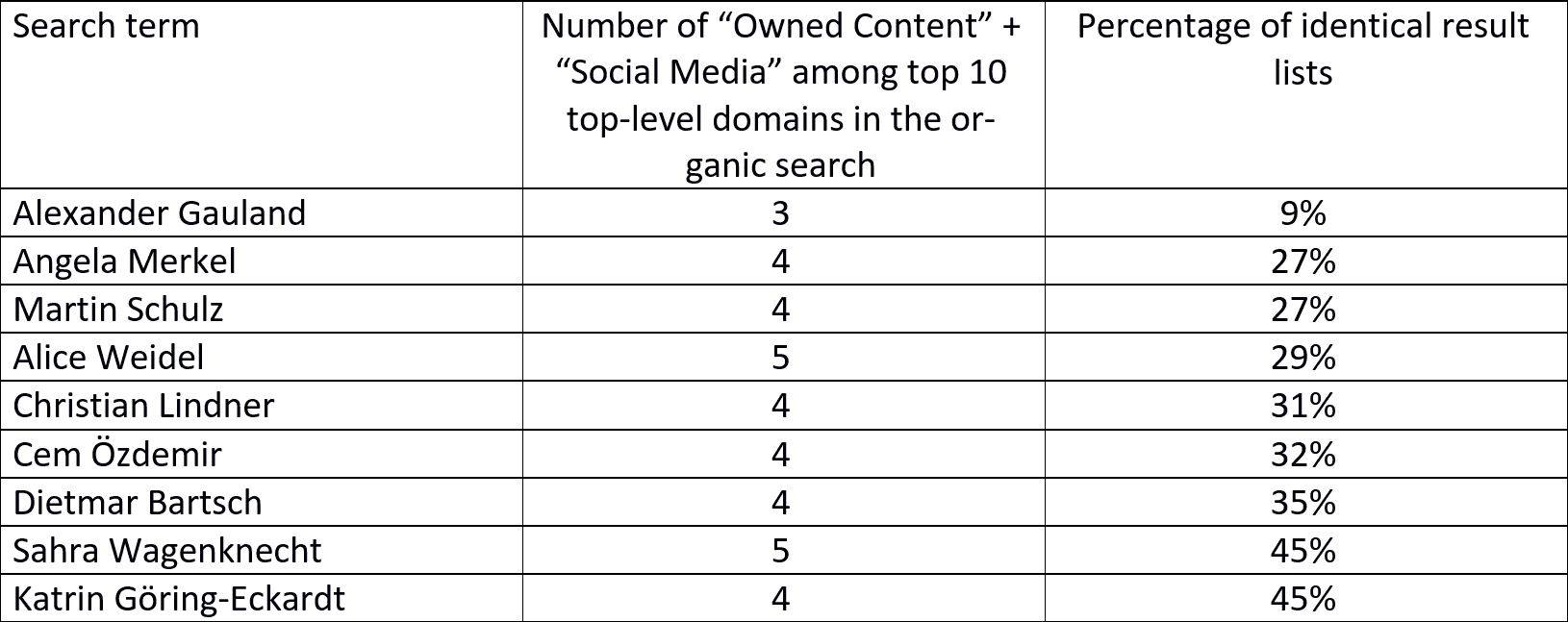}
    \caption{Comparison of the type of top-level domains of politicians among their top 10 top-level domains (Figure 9) and the percentage of identical result lists.}
\end{table}

In any case, it should be noted that even for people receiving the same links, the order is likely to be different since the proportion of those with exactly the same order of those with the same results is in most cases below 50\%. 
\subsection{Average number of common links}
Just because two people do not get exactly the same links does not mean that they get completely different links. Therefore, Table 9 and Table 10 each show the average number of common links for all pairs of data donors that look up the same search term at the same search time. For persons, the number ranges between 7.2 and 8.1 common links. This also has to be correlated to the average length of the search result lists. Since this is just over 9, on average only 1 to almost 2 links per randomly drawn pair of data donors are different. This is the "scope for personalization", which means the number of search results that could now be filled with links of various political content. In any case, the informational situation regarding the persons is -- on average! -- essentially the same. Therefore, on average we observe a very high overlap of the received news and information.\\

\begin{table}[ht!]
    \centering
    \includegraphics[width=0.5\textwidth]{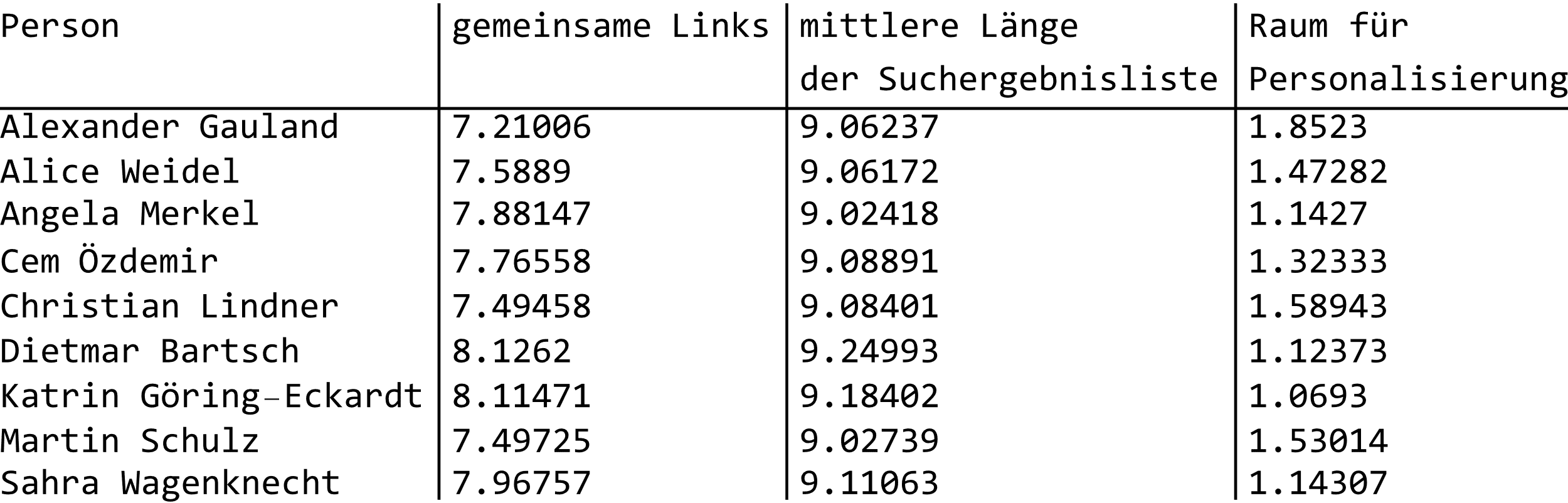}
    \caption{The mean result list length without top stories (3rd column) for the persons compared to the average number of URLs, which share a result list with all other result lists at the same search time (2nd column). From this, the average scope for personalization can be calculated by subtraction (4th column).}
\end{table}

For the parties, the result looks a bit different. Here, any two data donors searching for "Die Linke" at the same time have only 4.7 common links on average, for the "AfD" up to 7 links. Comparing this with the average lengths of the search result lists again, which fluctuate\footnote{The short organic search result lists are an effect of displayed top stories (the organic search result list usually contains 9 entries then) and can be further reduced by displaying deletion hints or the like. Especially with "Die Linke", such effects seem to have shortened the search result lists. Since the corresponding information was not saved on the search results page, we can not make a final assessment here.} between 8 and 9, for the parties an average of 2 (AfD), 3 (B{\"u}ndnis90/Die Gr{\"u}nen, Die Linke, FDP) and almost 4 (CDU, CSU) of non-shared links remain.\\

\begin{table}[ht!]
    \centering
    \includegraphics[width=0.5\textwidth]{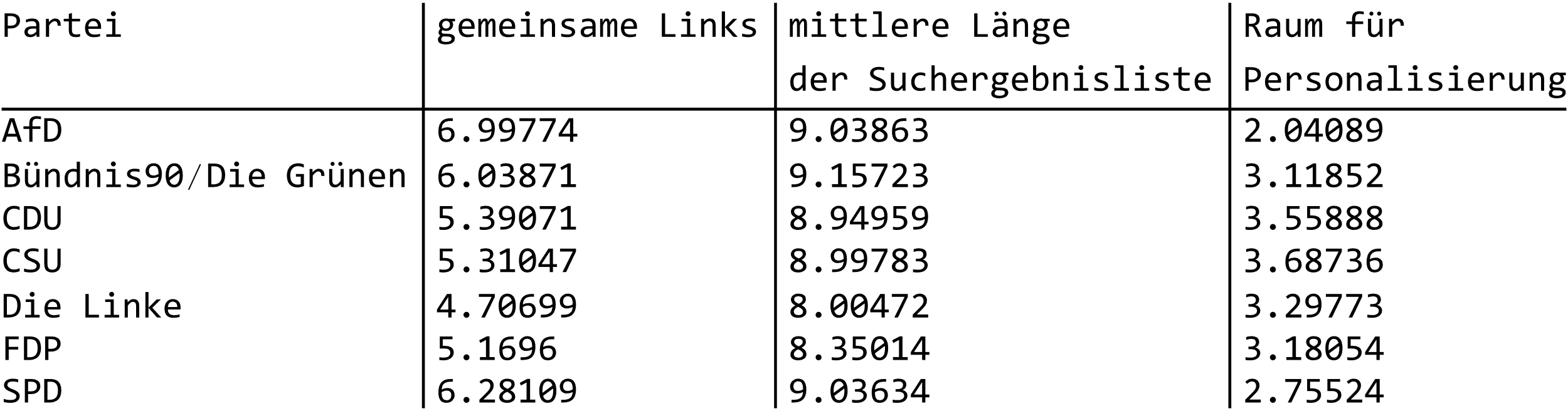}
    \caption{The mean result list length without top stories (3rd column) for the parties compared to the average number of URLs that share a result list with all other result lists at the same search time (2nd column). From this, the average scope for personalization can be calculated by subtraction (4th column).}
\end{table}

Figure 17 and Figure 18 show that the averages essentially originate from distributions that are skewed to the left. This is because each pair can share a maximum of 8 or 9 links, depending on the average length of the lists. Therefore, to the right this limit is closer than to the left. Only the AfD shows a bimodal distribution, e. g. 10\% of pairs of data donors sharing only 2 links. This finding will be further considered below (Section 5).\\

\begin{figure}[ht!]
    \centering
    \includegraphics[width=0.5\textwidth]{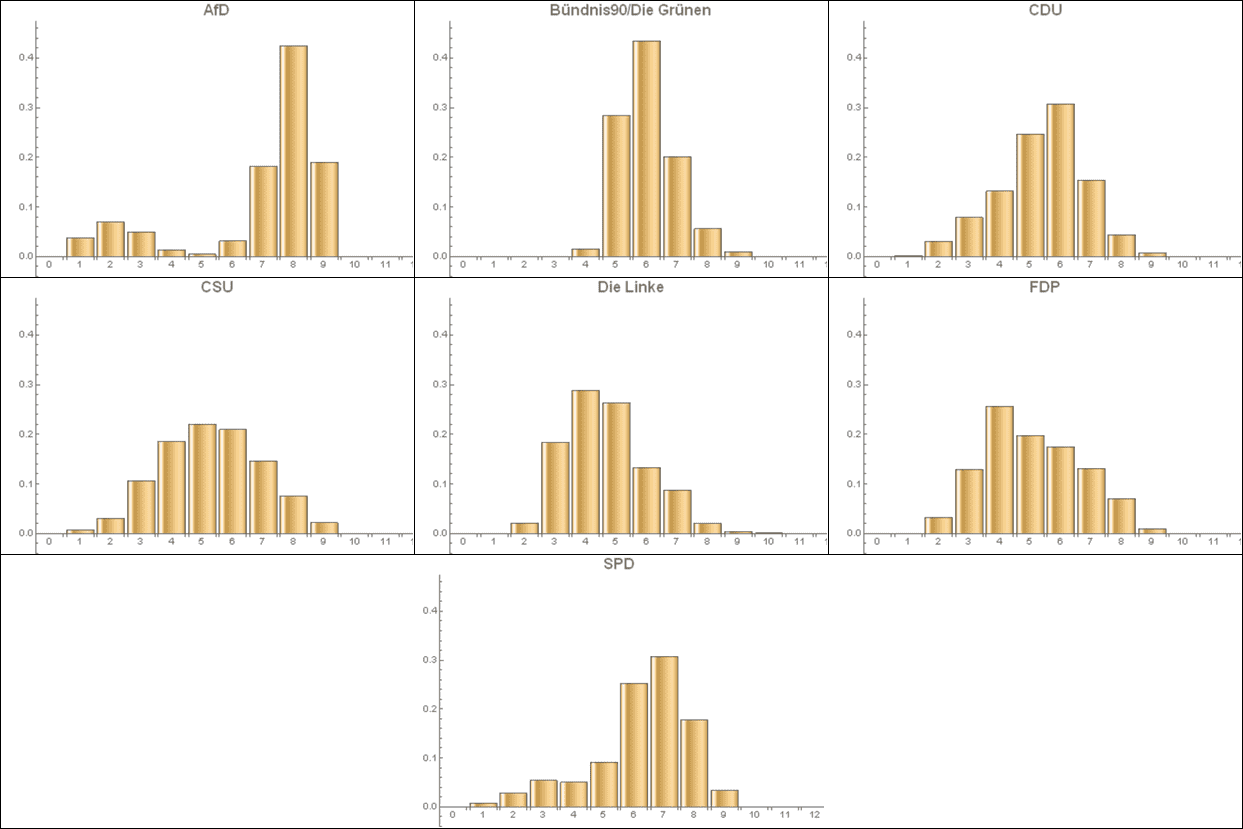}
    \caption{Distribution of the number of identical URLs in pairwise comparison of the organic search results for the parties.}
\end{figure}

\begin{figure}[ht!]
    \centering
    \includegraphics[width=0.5\textwidth]{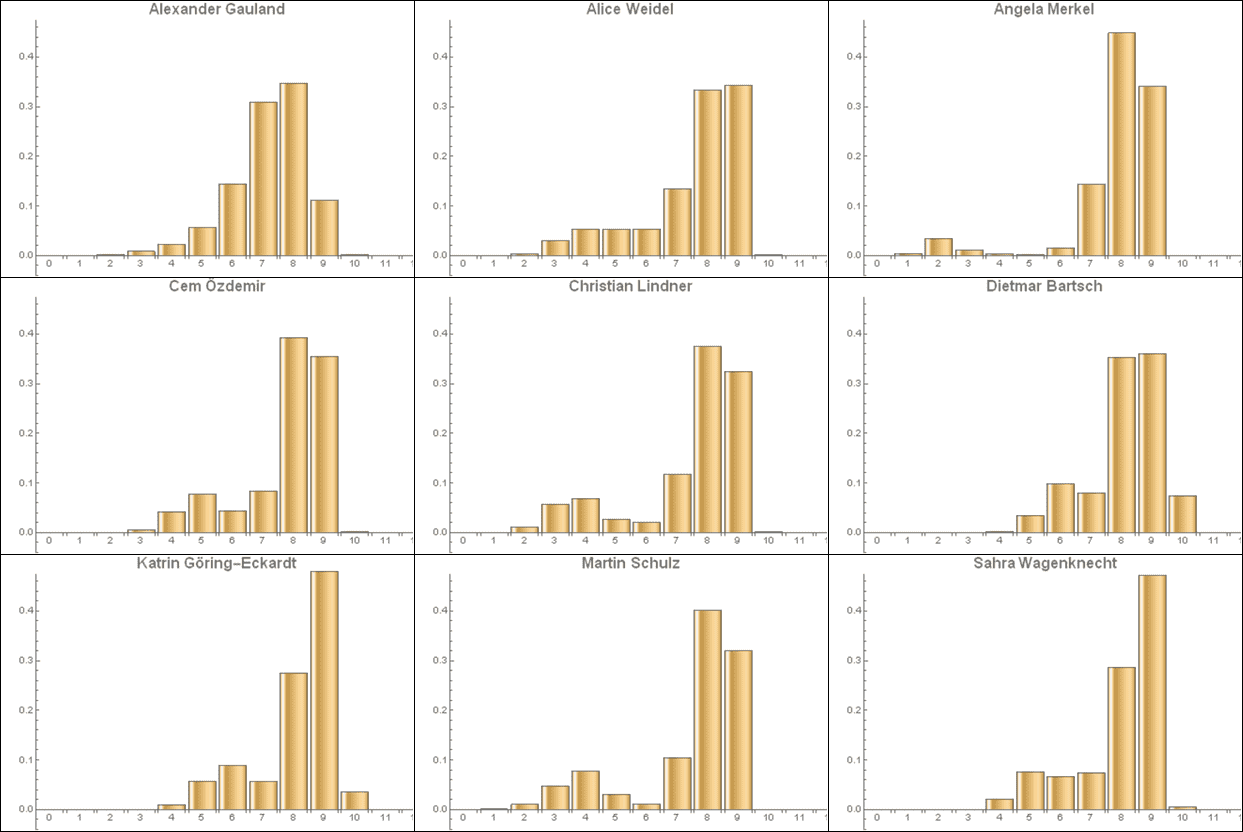}
    \caption{Distribution of the number of identical URLs in pairwise comparison of the organic search results for the persons.}
\end{figure}

We also conducted the same investigation on the Google News search result lists, the outcome of which we will discuss in the next section.
\subsection{Scope for personalization on Google News}
The search result lists on Google News almost always contain 20 entries, therefore the average rounded lengths are 20 for almost all of the search terms. Of these, for persons on average 17.3 to 18.3 are shared. The percentage of shared links is thus even higher than for the Google Search, leaving 2 to 3 links with possibly personalized messages.\\

\begin{figure}[ht!]
    \centering
    \includegraphics[width=0.5\textwidth]{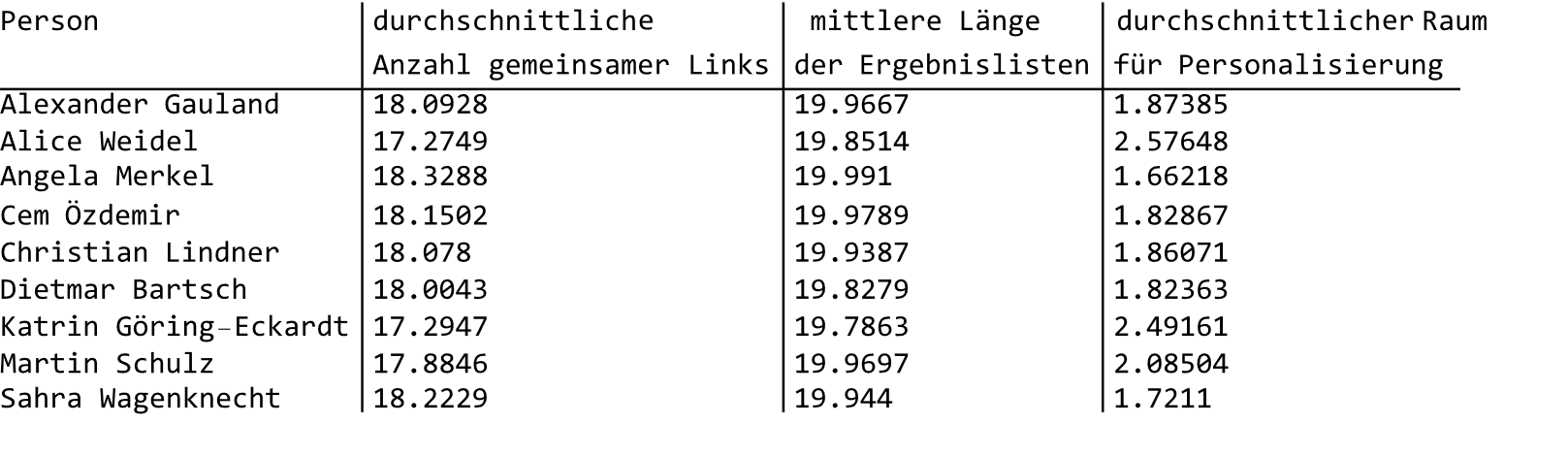}
    \caption{The mean result list length on Google News (3rd column) for the persons compared to the average number of URLs that share a result list with all other result lists at the same search time (2nd column). From this, the average scope  for personalization can be calculated by subtraction (4th column).}
\end{figure}

For the parties, the scope for personalized messages (on average) is also rather small and ranges from 1.3 to 2.1 for all but "B{\"u}ndnis90/Die Gr{\"u}nen", where up to 3 links from 18 could be personalized. It is important to note that only pairs of people are considered here. That which A and B do not share, B could share with C. Thus, these non-shared links do not necessarily have to be displayed to only one person, but may again affect smaller subgroups.

\begin{figure}[ht!]
    \centering
    \includegraphics[width=0.5\textwidth]{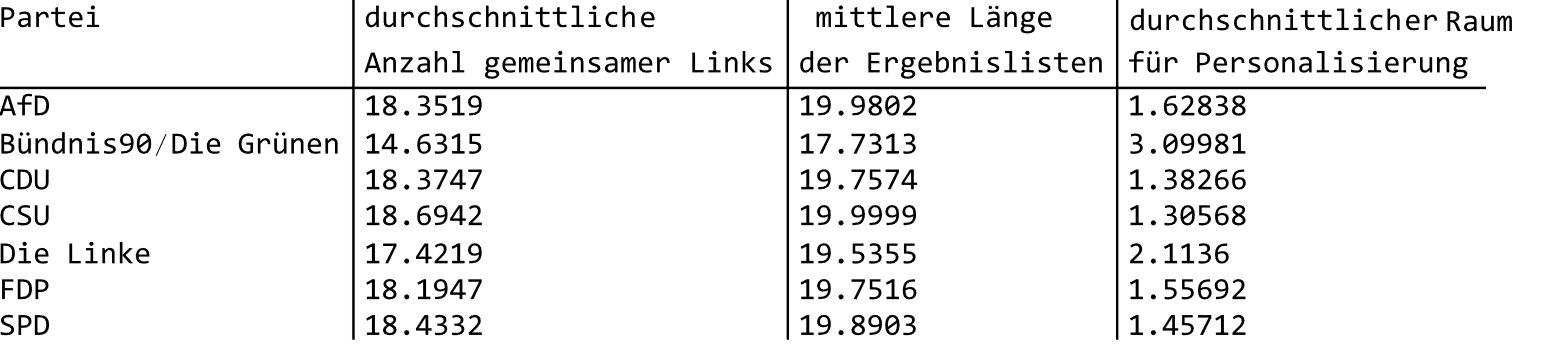}
    \caption{The mean result list length on Google News (3rd column) for the parties compared to the average number of URLs that share a result list with all other result lists at the same search time (2nd column). From this, the average space for personalization can be calculated by subtraction (4th column).}
\end{figure}

\section{Regionalization}
Due to the large amount of Owned Content, which can mainly be attributed to the many websites of regional branches or the websites of local politicians, regionalization of search results inevitably leads to many non-shared links. Therefore, we have once again categorized each party's Owned Content as regional/non-regional. We were rather conservative in our approach and have only marked those domains as regional which clearly referenced a city in the URL. For each pair of party search result lists, we first removed those regional URLs and then calculated the average number of non-shared links. Therefore, this represents a closer approximation to the possible number of personalized links, because the regional links are explicitly excluded. Table 11 shows the result of this refinement: While the data donors differ on average by 3 links for Die Gr{\"u}nen, only one of these is non-regional (and could be personalized). This is similar for the SPD, FDP, CDU, and Die Linke, each falling from (about) 3 non-shared links to below 2 non-regional links.\\

\begin{table}[ht!]
    \centering
    \includegraphics[width=0.5\textwidth]{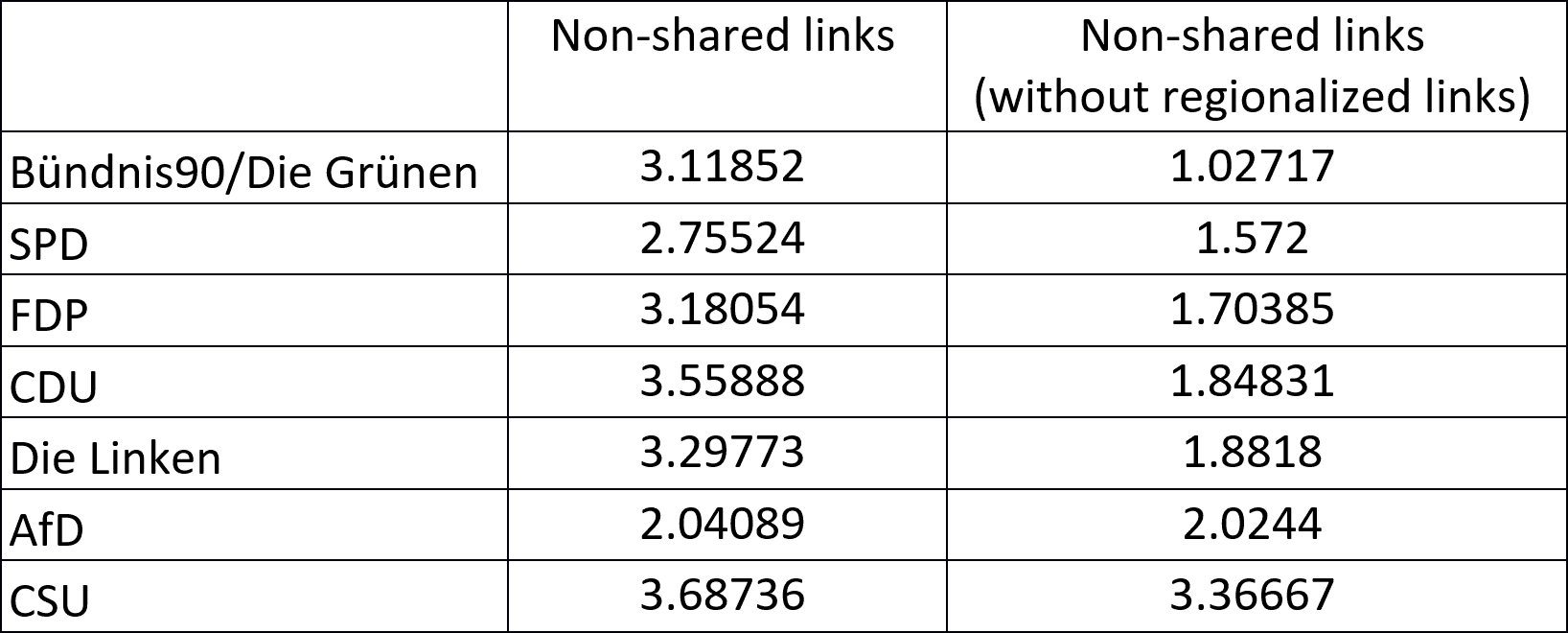}
    \caption{Refining the average number of non-shared links while taking into account regionalization. The left-hand column shows the average number of non-shared links for the entire search result list, and the right-hand one for the search result list after deleting the clearly regionalized websites (the respective local branches or individual local politicians).}
\end{table}

There is essentially no effect for the AfD to be seen -- yet that is not surprising, as only six top-level domains are displayed at all and no websites belonging to local branches were found. Of the six domains, four are of a regional nature (afd.berlin, afd.nrw, afd-bw.de, afd-fraktion-brandenburg.de). Interestingly,  little change occurs for the CSU, too. However, a look at the map of Germany showing the approximate locations of our data donors shows that Bavaria has fewer data donors who were still in the summer holidays during the first four weeks of the investigation period. Since most of the local branches are located in Bavaria, hardly any regionalized search results were delivered. A manual inspection of the search result lists shows that relatively many messages are delivered that overlap less strongly. Here, a more in-depth, content-related inspection appears sensible. Nonetheless, on average the search result lists clearly overlap more than half, so that here, too, a common information base is laid out.

\section{Dynamics}
An important subquestion of the project concerns the dynamic in which search results in the ordinary Google search change. One possible hypothesis here is that towards election day, Google's search results change in their similarity or the share of different media. In the sequence of the previous analysis, we first considered the dynamics of the respective share of top stories as the most prominently placed results (see Section 6.1). The second is an analysis of the number of delivered top-level domains per search time (see Section 6.2), followed by a look at the respective shares of editable URLs in the search queries, i. e. Owned Content, Social Media and (with some limitations) Wikipedia (see Section 6.3).
\subsection{Dynamics of the occurrence of top stories}
A first interesting point in the question of whether the search results have changed towards the election was whether and when top stories are displayed and whether the proportion of search results with top stories has increased towards the end of the election campaign, for example. In this case, effects such as a higher news density could lead, as an example, to Google delivering more search result lists with top stories shortly before the election than five weeks before that.\\
However, the analysis of the occurrence of top stories has shown that across all search terms and search times, the proportion hardly fluctuates. Percentages always range in a corridor around 90\% with up and down swings of no more than 10\% (see Figure 21). Only when looking at the individual search terms stronger deviations have been found.\\

\begin{figure}[ht!]
    \centering
    \includegraphics[width=0.4\textwidth]{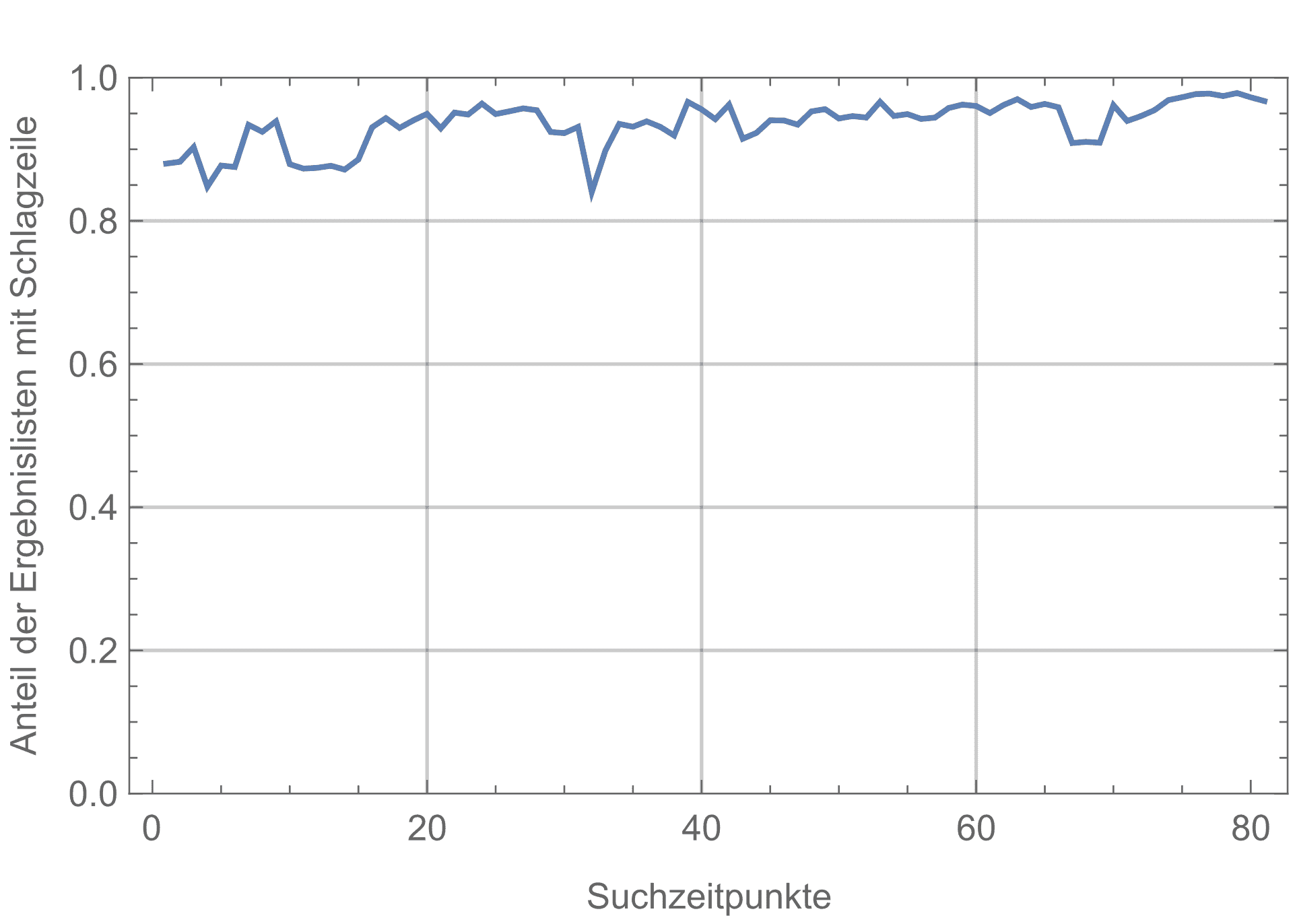}
    \caption{Proportion of all result lists for a search time, where at least one top story was delivered.}
\end{figure}

There are clear outliers for Katrin G{\"u}ring-Eckardt and Dietmar Bartsch (see Figure 22). In contrast, the shares of search result lists with top stories for the search terms CSU, SPD, Angela Merkel, Martin Schulz and AfD are very stable, although at the beginning of the investigation period their proportion of top stories (search time 6 to 22) is still subject to fluctuations (see Figure 23). The isolated, complete absence of top stories (see Figure 22) may indicate that it is decided by Google whether or not to deliver a top story according to the latest news -- but these deviations can not be fully explained from the outside. Without further information it is particularly inexplicable why, for the same search time and search request, Google delivers top stories for some participants and does not do so for others. Especially the very stable values of some search terms (CSU, SPD, Angela Merkel and Martin Schulz) contradict a browser setting of these users which would suppress such a top story in any form, because then these users would not have received top stories for other search terms (such as CSU, SPD, Angela Merkel). However, these search terms show many search times where all users received top stories.\\

\begin{figure}[ht!]
    \centering
    \includegraphics[width=0.5\textwidth]{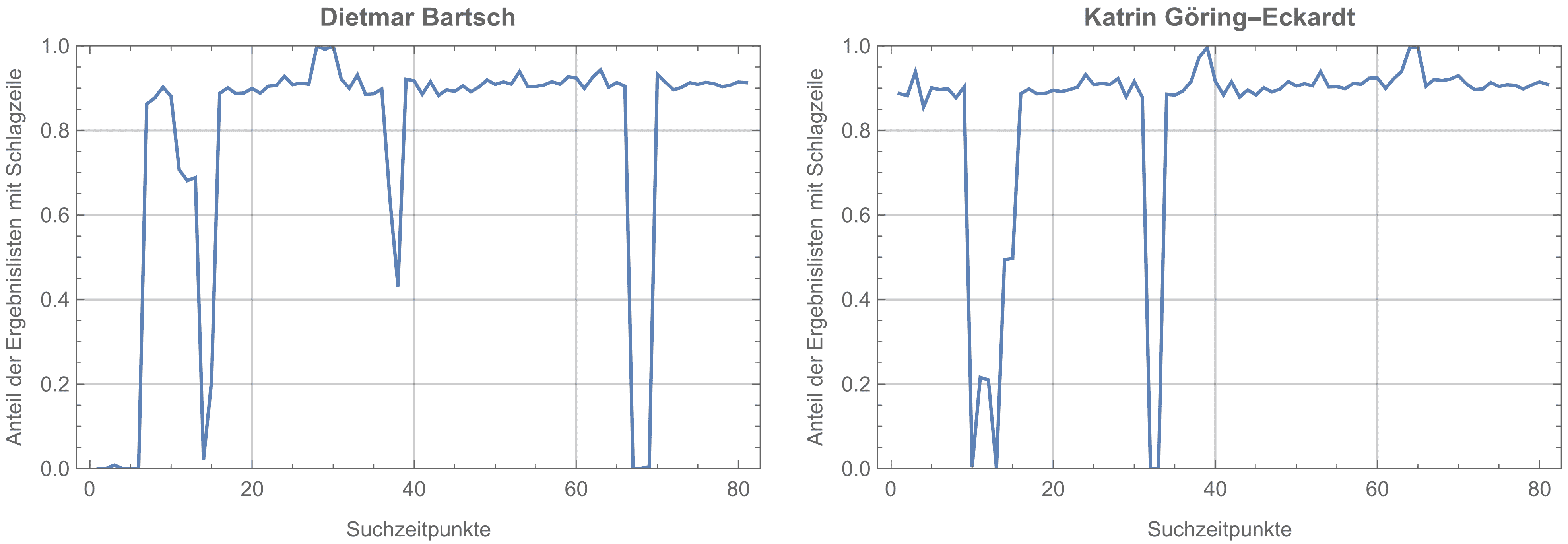}
    \caption{Proportion of result lists with top stories for the search terms "Dietmar Bartsch" and "Katrin G{\"o}ring-Eckardt" show clear deviations from all others.}
\end{figure}

This inconsistency also becomes clear when looking at individual users. For example, with the same user and search term sometimes top stories appear and sometimes they do not (see Figure 24). Overall, however, the proportion of search result lists with top stories is relatively stable and shows no particular temporal patterns with regard to the federal election.

\begin{figure}[ht!]
    \centering
    \includegraphics[width=0.5\textwidth]{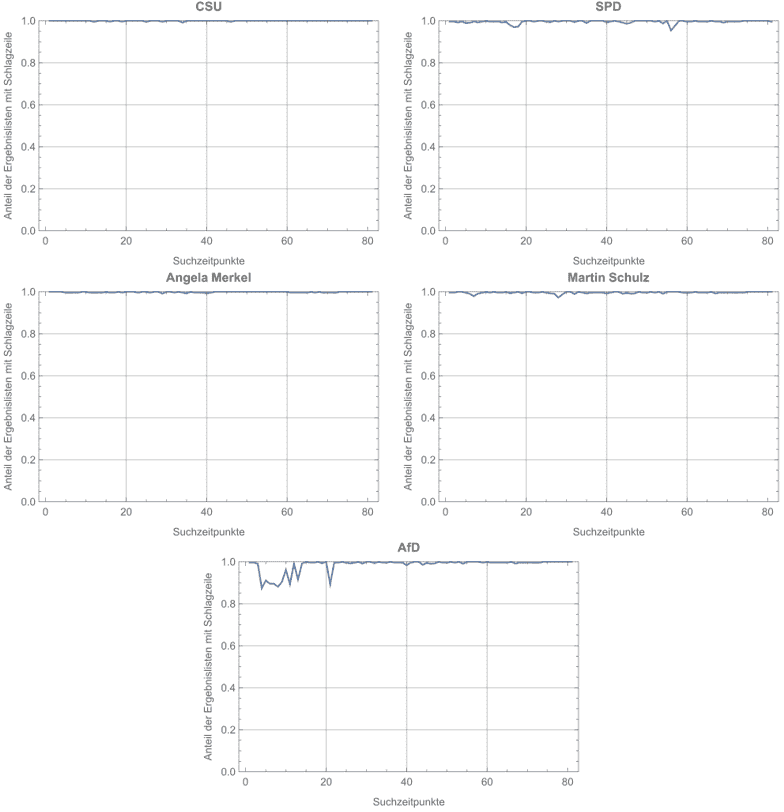}
    \caption{Proportion of all result lists that contain top stories, for the search terms CSU, SPD, Angela Merkel, Martin Schulz and AfD.}
\end{figure}

\begin{figure}[ht!]
    \centering
    \includegraphics[width=0.4\textwidth]{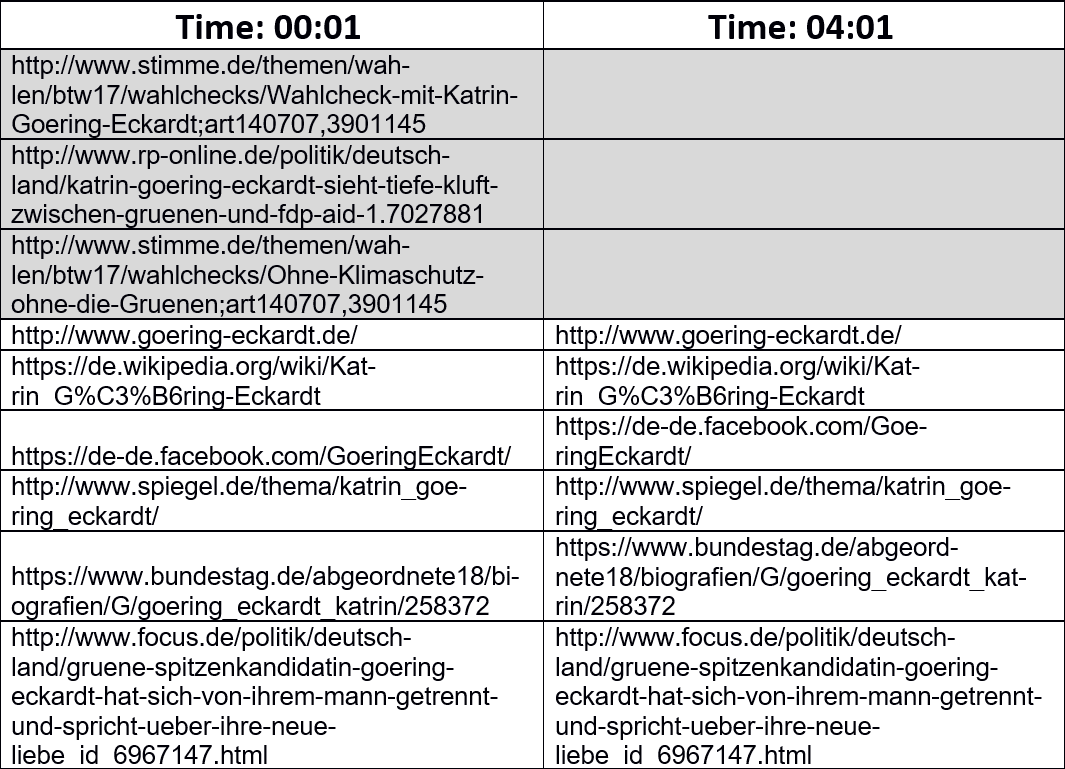}
    \caption{Search results of a user from August 28th 2017 for the mentioned times and the search term "Katrin G{\"o}ring-Eckardt". Top stories were once displayed to the same user at different, consecutive search times (highlighted in gray) and then they were not.}
\end{figure}

\subsection{Dynamics of the number of different top-level domains}
In processing of the number of different top-level domains that were delivered to at least one data donor at a search time, on the one hand already recorded trends (see Figure 25, 26) confirm that search result lists for parties generally have more top-level domains than results lists for persons. On the other hand, this number fluctuates quite strongly from party to party: While only about 22 top-level domains characterize the results on the AfD at most search times, the number is  around 180 for Die Gr{\"u}nen (with large deviations). Essentially, the number of top-level domains of the parties is reflected here again, and the fluctuating extent of regionalization is thus visible.\\
For the persons the number is between 17 and 18, which also shows that the number of shared links (including top-level domains) is very high for the persons.\\
Although the fluctuations over the search period are sometimes very pronounced, there are no striking temporal patterns in the number of different top-level domains.\\

\begin{figure}[ht!]
    \centering
    \includegraphics[width=0.5\textwidth]{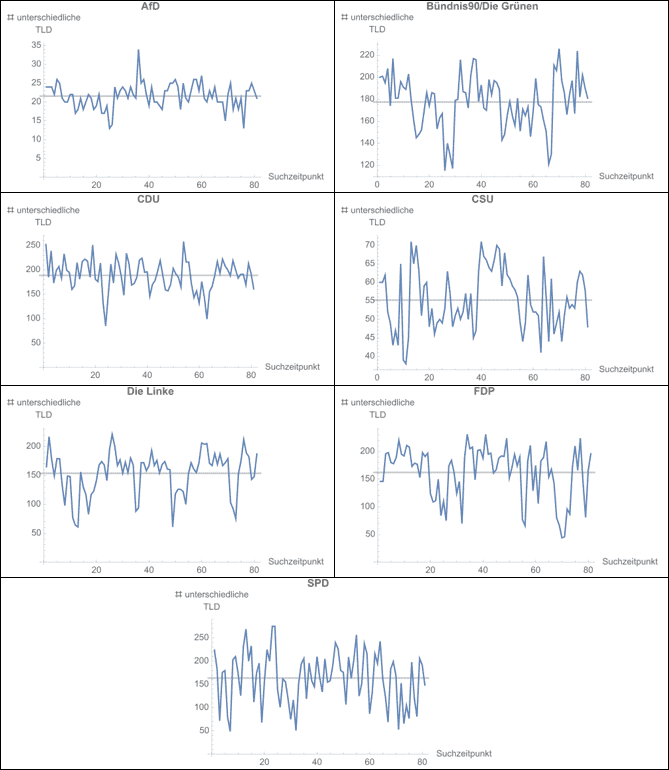}
    \caption{Number of different top-level domains over the 81 search times for the Google search of parties.}
\end{figure}

\begin{figure}[ht!]
    \centering
    \includegraphics[width=0.5\textwidth]{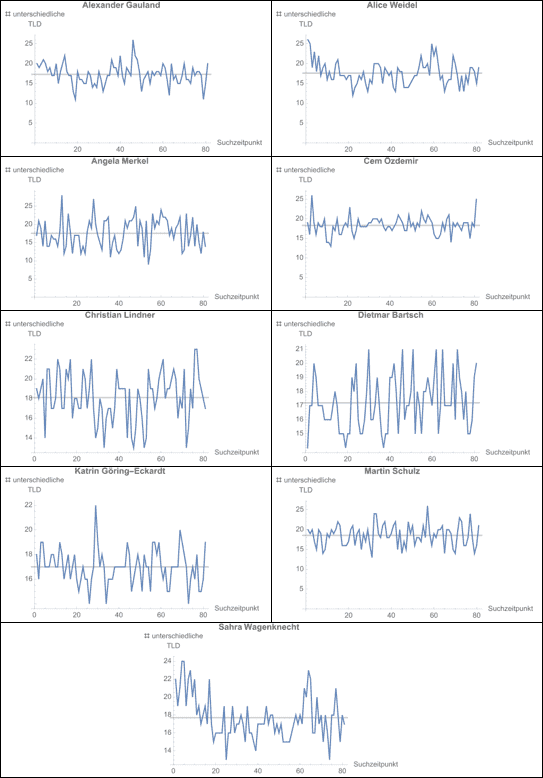}
    \caption{Number of different top-level domains over the 81 search times for the Google search of persons.}
\end{figure}

\subsection{Dynamics of proportions of editable contents}
Over time, does the proportion of URLs that are on top-level domains and that are editable by the parties or persons (in principle) change over time? We also examined this question for all search times during the investigation period. There is a slight drop for all parties (except Die Gr{\"u}nen) for the search times right before the election. It should be noted that the last 6 search times are from the election weekend. Presumably this is caused by an increased share of news, although this only takes effect in the last days before the election.\\

\begin{figure}[ht!]
    \centering
    \includegraphics[width=0.5\textwidth]{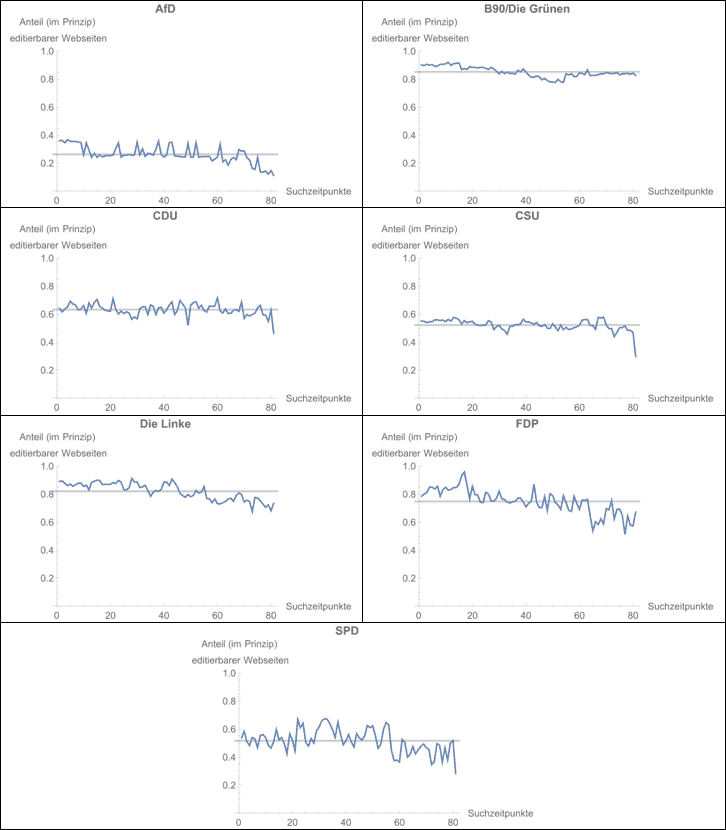}
    \caption{Proportion of (in principle) editable websites for the respective party of all result links of the respective party. Here, the proportion for each of the 81 search times is applied and the mean value is drawn in in gray.}
\end{figure}

For most persons, too, this short-term decline in editable websites can be seen among the search results, which in some cases already starts one week before the election (Gauland, Lindner, {\"O}zdemir and Wagenknecht). The result was not pursued further, because this report deals with the key issues of personalization, regionalization and filter bubbles. However, for media scientists it could prove to be an interesting initial observation that can be followed up on by means of the publicly available data from the data donation.

\begin{figure}[ht!]
    \centering
    \includegraphics[width=0.5\textwidth]{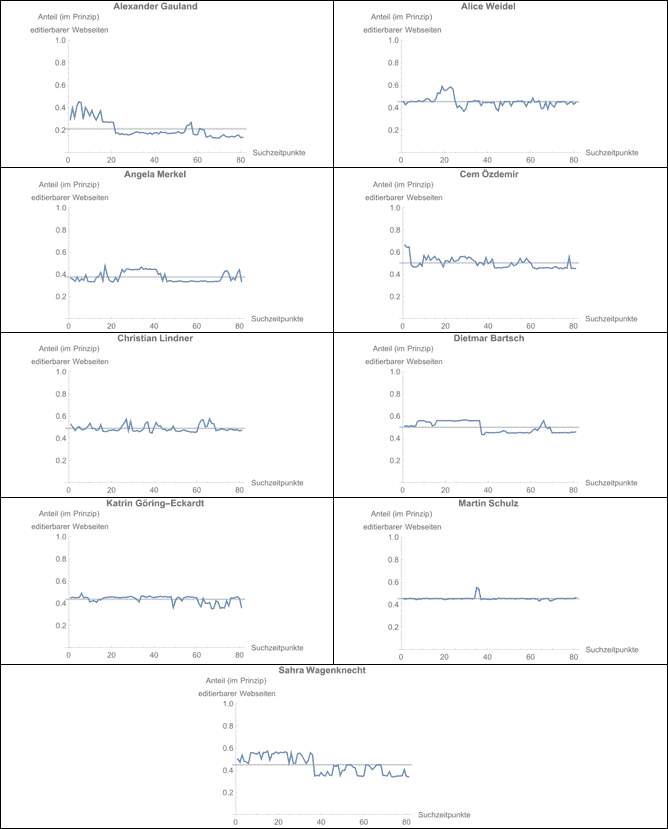}
    \caption{Proportion of (in principle) editable websites for the respective persons of all result links for the respective person. Here, the proportion for each of the 81 search times is applied and the mean value is drawn in in gray.}
\end{figure}

\section{Search result lists with low similarity to all others}
Even though most of the search result lists are very similar on average, some pairs of search result lists beyond the average showed almost no overlap. As previously mentioned in the first interim report, a manual inspection of search result lists showed the occasional result list that differs greatly from all of the others\footnote{Section 4 of the first interim report (Krafft et al., 2017)}. It was determined that these lists often contained links to websites in other languages (mostly English). It is difficult to conclusively cluster and identify these links automatically, partly because the link itself does not necessarily give any insight on the language of the website it links to, e. g. some websites end in ".com" and  still refer to a German website like "handelsblatt.com". Most social media pages also end in ".com" (twitter.com, facebook.com) and the language displayed depends on the user being logged in. On Wikipedia pages the language is encoded into the first two letters of the URL, so "de.wikipedia.org" refers to the German, and "en.wikipedia.org" to the English version of that page. Furthermore it is unclear if all deviant clusters have to contain at least one website in a foreign language.\\
We also followed a second approach, that also requires manual preparation of the data: First links were removed that were  seen by least 70\% of all users for each search term and time, since their popularity suggests a lack of personalization. In the next step, we clustered the lists with the remaining links, so that the lists within these clusters are similar once more. The original lists of these groups were then compared to every user outside of the group, to find the number of common links. For example: We find 5 users, who retained only 4 entries in their lists, after the popular links for the search term "Die Linke" were removed, but still shared at least 3 links with each other -- they form a cluster. For these 5 users, we compared their original result lists with every other user outside of the cluster to find the average number of common links. If this average falls below 3.5 the cluster is considered to be a very distinct group. During a manual inspection of the identified distinct clusters we found that, especially for the search terms "Die Linke" and "B{\"u}ndnis90/Die Gr{\"u}nen", most of these clusters were regional groupings, whose search result lists differed only in the large selection of local branches of the party.\\
In summary we can say that automatically identifying groups of users that receive different content according to the filter bubble theory (not originating from regionalization), has proven to be quite a complex task.\\
A third approach showed a clear pattern that does indicate a stable filter bubble, which would have to be subject to further studies: Based on the time of publication that accompanies every top story, it is possible to detect the main language Google assigns to a user. Our data set contains publication times in Norwegian, English, German and French. This made it possible to partition the search result lists. In all examined cases, we saw a separation of users by Google into "presumably German" with high similarity of results per search term and time, and "presumably foreigners" who shared unusual results which were then mixed with results of the suspected, preferred language.\\
This was done via manual inspection, since there was no possibility to automatically separate users into those that are considered German-speaking by Google and those that are not (primarily) German-speaking. Some users received German publication times once and only English publication times after that, and many of the users manually flagged as having partially non-German search result lists did not receive any top stories and therefore did not receive publication times.\\
This pattern will be demonstrated with some selected examples below.
\subsection{Example: Search result lists from a French IP}

\begin{figure}[ht!]
    \centering
    \includegraphics[width=0.5\textwidth]{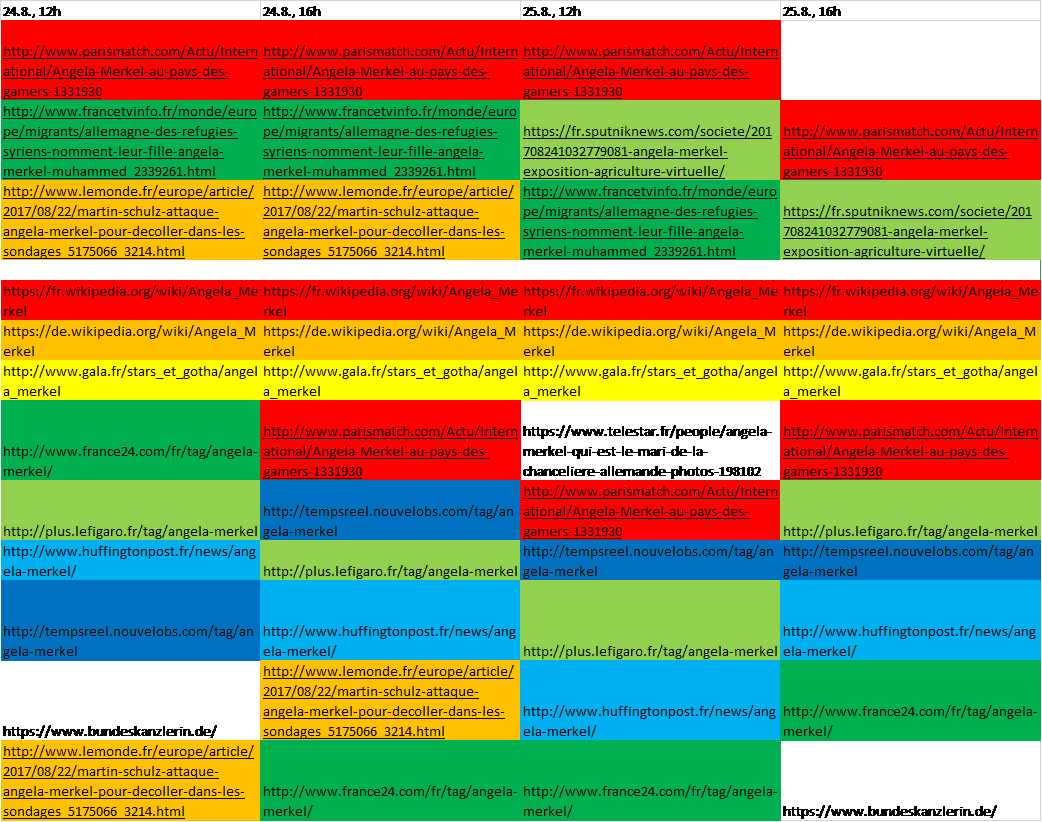}
    \caption{Search results for the search term "Angela Merkel" of a user with an IP address in France, close to Paris. The results are divided into three top stories and the organic search results below.}
\end{figure}

Figure 29 shows the search results of a user in France. For the result lists containing top stories one can clearly see that Google suspects a French person, since the publication times that Google adds to every news entry are French (e.g. "Il y a 2 heurs"). The top stories themselves link to French websites, which discuss the German election. It is interesting to see that some lists contain German results. The results for the search term "Angela Merkel" overall look like a sensible selection of websites -- at least for a person living in France.\\
For the search term "AfD" the same person receives -- along with results concerning the German political party -- websites from French initiatives with the same acronym, mainly  L'Agence fran{\c{c}}aise de d{\'e}veloppement, a French organization fighting diabetes and supporting people with autism and their families. Following the same method of looking for international institutions or organizations with the same acronym as the political parties we found the French Cultures et Soci{\'e}t{\'e}s Urbaines and the American universities Charles Sturt University and Colorado State University while searching for "CSU".\\
The Google search for the German politicians is for the most part without error, meaning that the user almost always receives information about the persons searched for: The results contain mostly the social media profiles and -- depending on the prominence of the politician -- their French Wikipedia entry and French news outlets; this is especially the case with Angela Merkel and Martin Schulz. Apart from that, the user receives German news websites and social media pages. An exception to this is Cem {\"O}zdemir: One Twitter profile received belongs to another person with the same name\footnote{https://twitter.com/esekherif\_} and results concerning a football player on the transfer market\footnote{https://www.transfermarkt.fr/cem-ozdemir/profil/spieler/170442}, both of which are not members of "Die Gr{\"u}nen".\\
The occurrence of links in the (suspected) native language of users outside of Germany is without question useful and the amount of German-speaking websites received by the examined user probably originates in the lack of French-speaking websites for these particular search terms. It is not possible to further determine how the choice was made whether and when a user in a foreign country receives links to German websites.\\
It is a lucky coincidence that we received data from multiple data donors, all with an IP address originating in Kaiserslautern, but in different languages, namely English, Russian and German. This allows us to eliminate the influence of regionalization based on the IP address.\\
In the next step we examine the search results of a user receiving links to Russian websites from Google while searching from a German IP address from Kaiserslautern.
\subsection{Example 2: Google search from Germany with partially Russian results}
Our manual inspection showed a person located in Kaiserslautern according to their IP address, who often received Russian Wikipedia pages while searching for politicians and political parties. Our browser plug-in recorded the IP address as well as the keyboard language used, which was set to Russian for this person. Table 12 shows a typical search result list for this user when searching for "Angela Merkel".\\

\begin{table}[ht!]
    \centering
    \includegraphics[width=0.5\textwidth]{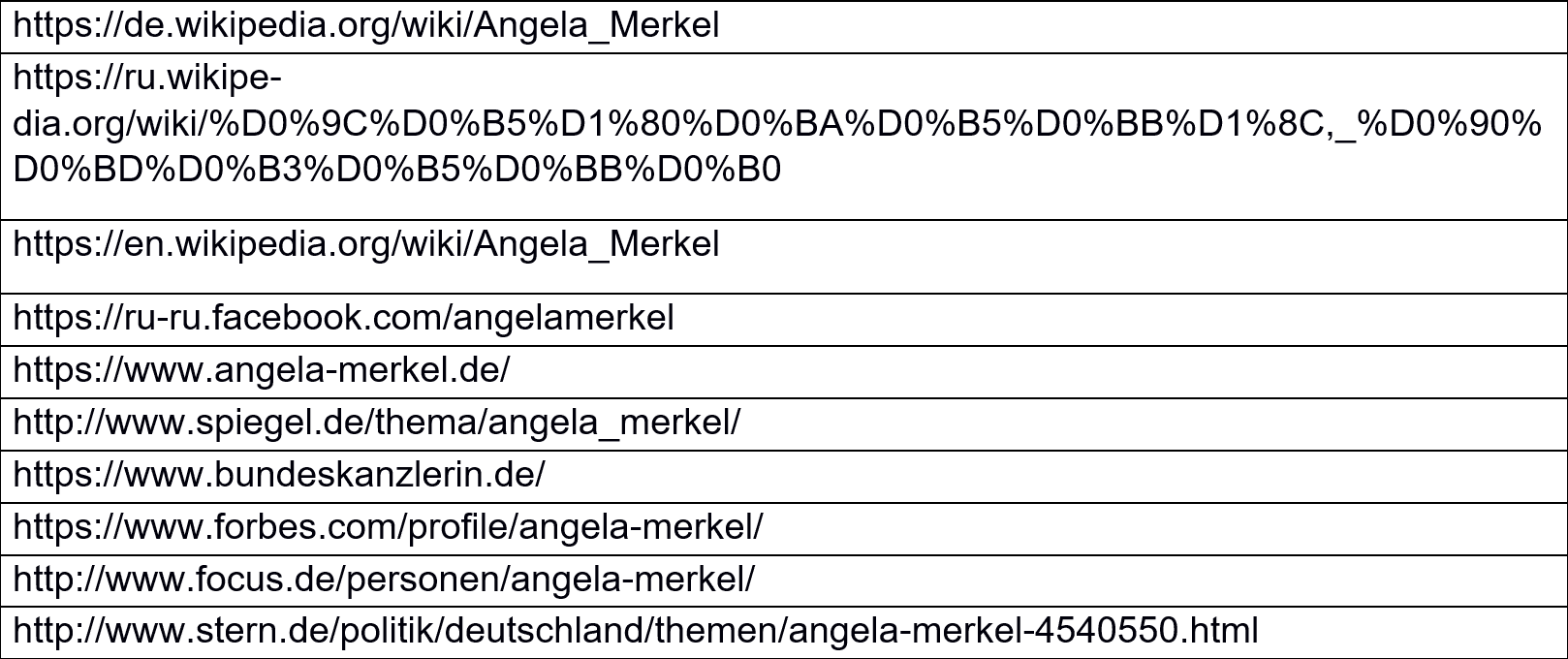}
    \caption{Typical search result list for the search term "Angela Merkel" for a user who donated around 40 result lists, sometimes containing Russian results.}
\end{table}

The second link of this result list refers to a Russian Wikipedia page of Angela Merkel, the third link to the English version. The fourth refers to the Facebook page of Angela Merkel, which is predominantly German content-wise. If someone were to open that page while logged into Facebook and with his or her preferred language set to Russian, the text supplied by Facebook would be in Russian (e.g. the user display, etc.). Thus, this would  not change the content of the politician. Lastly a second English-speaking website (forbes.com) appears near the end of the result list.\\
Therefore the user did not receive any Russian websites, but Google seems to regard the Russian Wikipedia pages as appropriate. It is unclear why this is the case. We observed that for a lot of users English keyboard settings do not necessarily imply English results, which is why we do not think the Russian keyboard settings are the (only) reason for these results. It is noteworthy that this person was always logged into Google, which could be the source of the information.\\
After that we compared the results received by the person from Kaiserslautern, who received some Russian results, to other users from Kaiserslautern and another user outside of Germany, namely the person mentioned before from France (Figure 30).\\

\begin{figure}[ht!]
    \centering
    \includegraphics[width=0.5\textwidth]{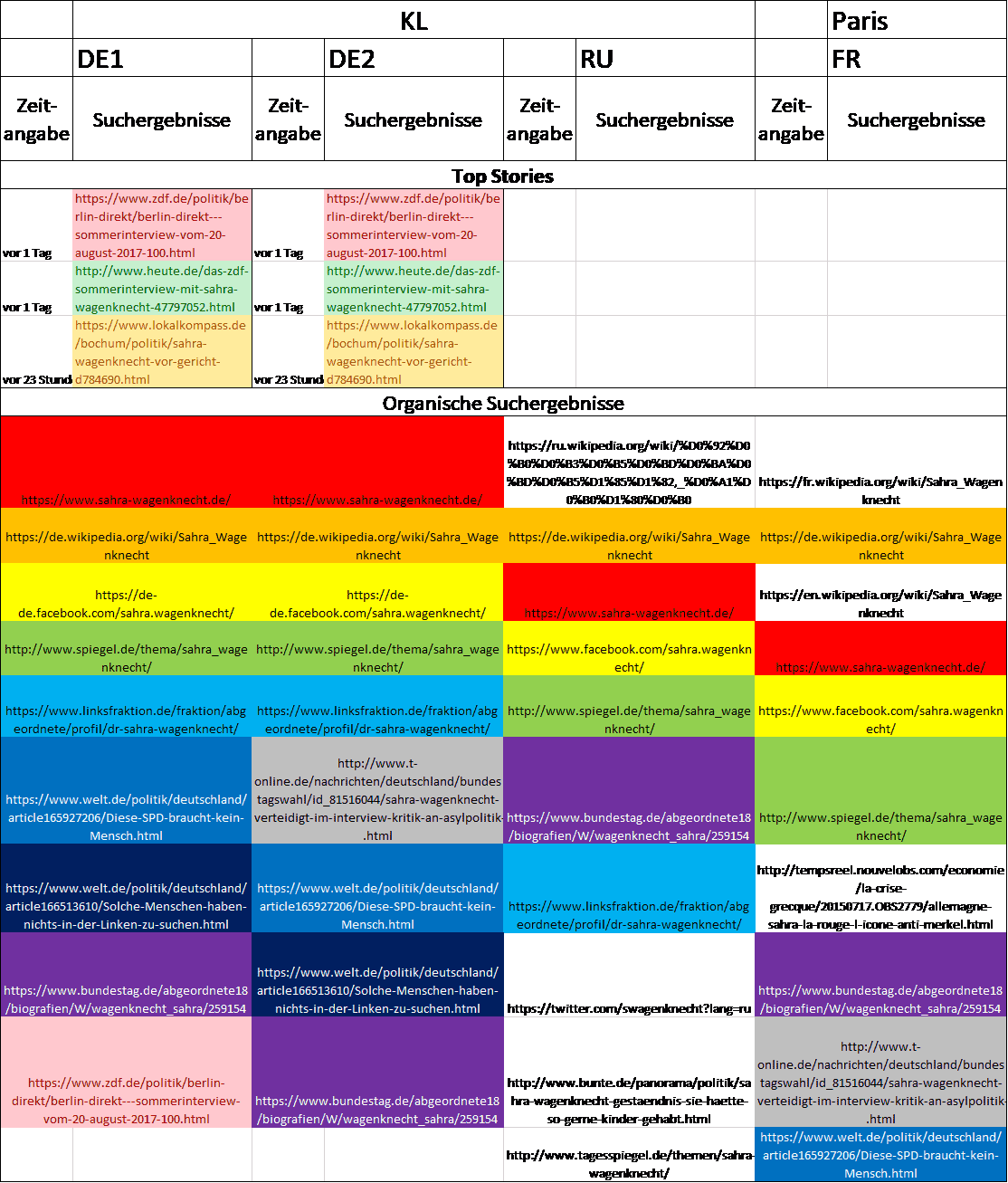}
    \caption{Search result lists for the search term "Sahra Wagenknecht" on August 22, 2017 from users with IP addresses in Paris (rightmost column) and Kaiserslautern. Identical results are represented by the same color. The first two columns show users whose result lists are constantly German. The third column shows the search results of a person who sometimes received Russian results. The fourth column shows the results of a user from Paris, France according to their IP address.}
\end{figure}

It is noteworthy that users who received non-German results from Google did not  receive any top stories. The user from Kaiserslautern with a link to the Russian Wikipedia website of Sahra Wagenknecht also received a link to her Twitter profile. The last part of this link "lang=ru" only affects the content supplied by Twitter (e.g. translations of the navigational buttons like "Messages", "Follow/ Unfollow") -- the tweets themselves are in German. The user also receives links to an overview page from the Tagesspiegel and an article from Bunte, both concerning Sahra Wagenknecht. The user from France received an article about "Sahra la Rouge" and the English Wikipedia entry.\\
Interestingly the user from Kaiserslautern does not receive a Russian Wikipedia entry for every person that is present there: Except for Dietmar Bartsch and Katrin G{\"u}ring-Eckardt, every politician included in our set of search terms is represented on the Russian Wikipedia website, but not always displayed to the user.\\

\begin{figure}[ht!]
    \centering
    \includegraphics[width=0.5\textwidth]{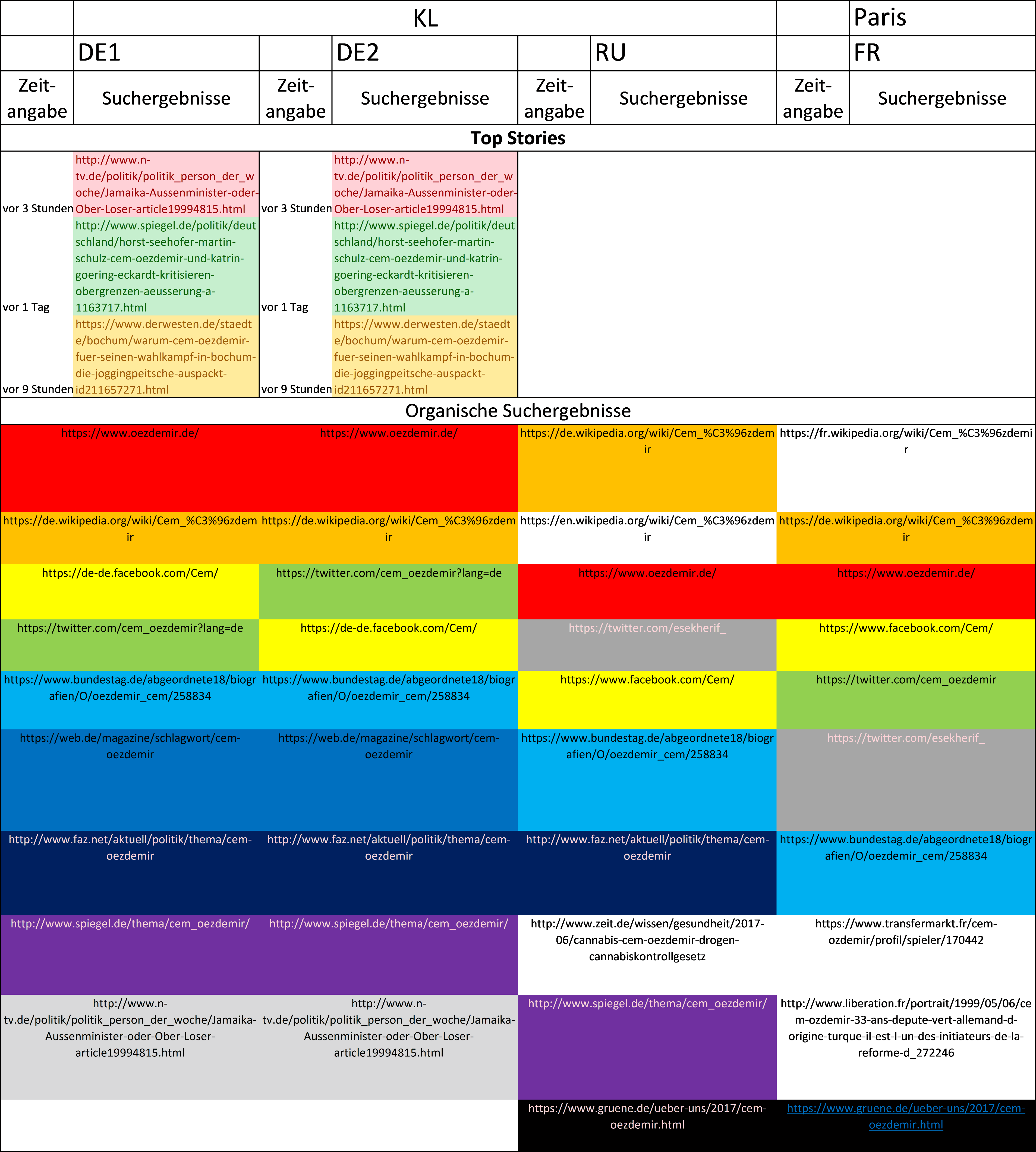}
    \caption{Search result lists for the search term "Cem {\"O}zdemir" on August 22, 2017 from users with IP addresses in Paris (rightmost column) and Kaiserslautern. Identical results are represented by the same color. The first two columns show users whose result lists are constantly German. The third column shows the search results of a person who sometimes received Russian results. The fourth column shows the results of a user from Paris, France according to their IP address.}
\end{figure}

Figure 31 shows the search results of 4 users for the search term "Cem {\"O}zdemir" and for the same search time. The user from Kaiserslautern receives the German and English Wikipedia entry, the French user gets the German and French entry. Apart from the links that they share with the users receiving constantly German result lists, they also share the Twitter profile of another person with the name "Cem {\"O}zdemir" who is not the German politician. The same is true for their respective last entry, a link to a sub-page of the top-level domain "www.gruene.de" that is rarely seen elsewhere. In summary both of the users who received non-German results from Google share 6 links. The French user also received two French websites along with the reference to the football player and the correct Twitter profile of Cem {\"O}zdemir. The result list of the user from Kaiserslautern, who also received Russian results, contained a news article from "Zeit" which was not received by the other three users. He or she shares 7 and 6 links respectively with the other users from Kaiserslautern.\\
A lot more frequent than a person who receives Russian Wikipedia websites are users with IP addresses in Germany and other countries, who receive mixed language results, consisting of English and German websites. Among these is a user with an IP address from Kaiserslautern. We therefore looked at four result lists of users with an IP address based in Kaiserslautern for all political parties at the same search time. Two of those users received only German results (DE1 and DE2), the user with partially Russian results (RU) and a user with partially English websites originating in the USA (USA). Figure 32 shows the number of common results for each party.\\

\begin{figure}[ht!]
    \centering
    \includegraphics[width=0.4\textwidth]{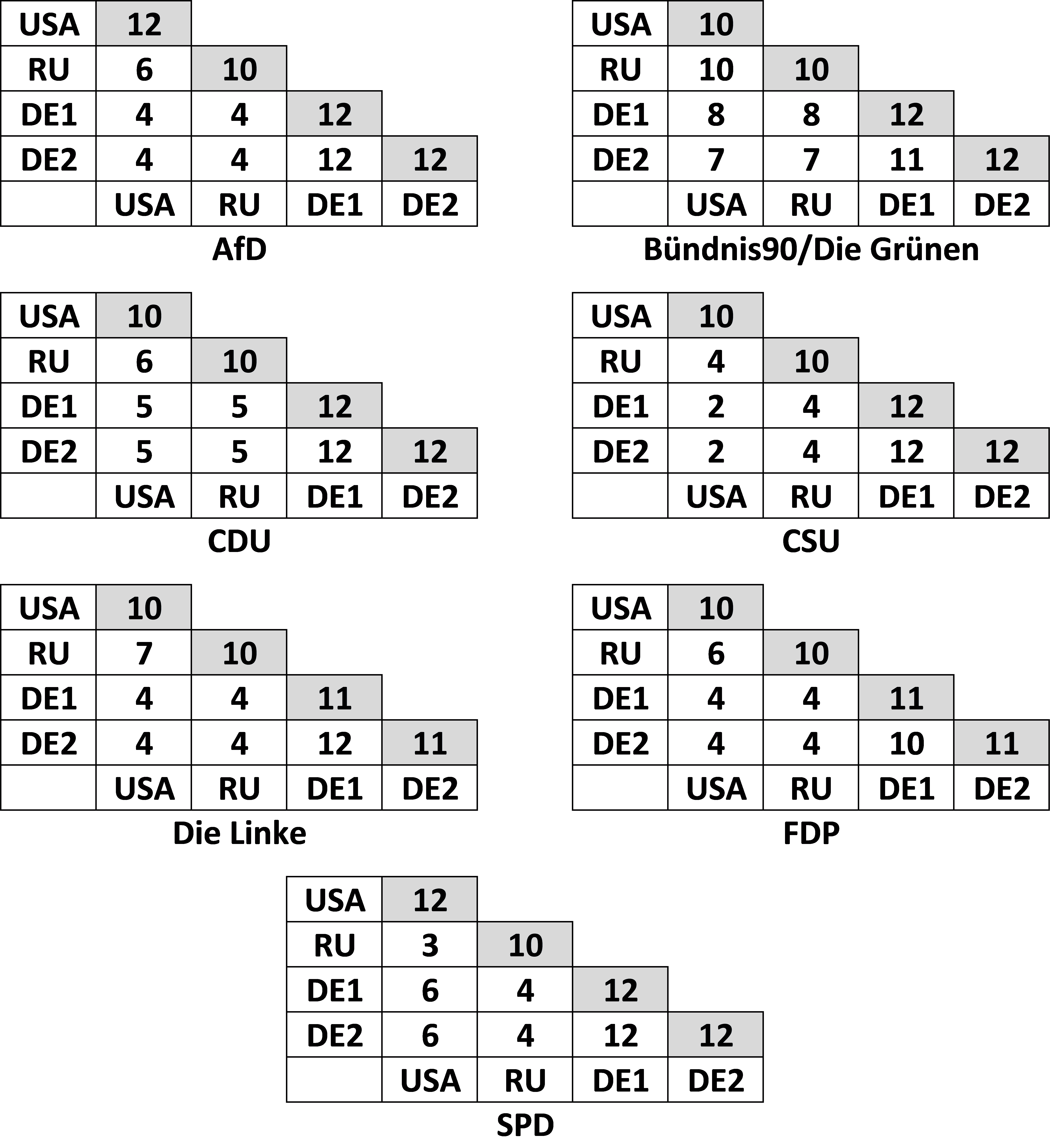}
    \caption{The result lists of four users with IP addresses based in Kaiserslautern for all parties on September 5, 2017 were compared to find the number of common links. Two users only received German results (DE1 and DE2), one user also received Russian and English websites (RU) and one user websites with American origin (USA). Displayed are the number of identical links for each pair -- the diagonal fields (gray) show the total number of links contained in the person's search results.}
\end{figure}

It is noticeable that the two users with purely German results almost always have identical search result lists. A manual inspection shows that even the order of the results is mostly identical. Surprisingly, the remaining users have more links in common for five of the parties and the same amount of links in common for two parties as the purely German users, even though they searched from the same location based on IP addresses.\\
To further enhance this pattern we did a last comparison to look at all search result lists from Kaiserslautern and all result lists outside of Germany based on IP locations for one search term and time (September 22, 2017, "Angela Merkel"). For every person except one top stories were included, so we could determine whether Google supplied them with English or German publication times ("1 hour ago" vs. "Vor 1 Stunde"); other languages were not represented in this sample. The astonishing result is shown in Figure 33: The similarity between the search results is highly dependent on whether Google German-speaking or an English-speaking Person.\\

\begin{figure*}[ht!]
    \centering
    \includegraphics[width=0.9\textwidth]{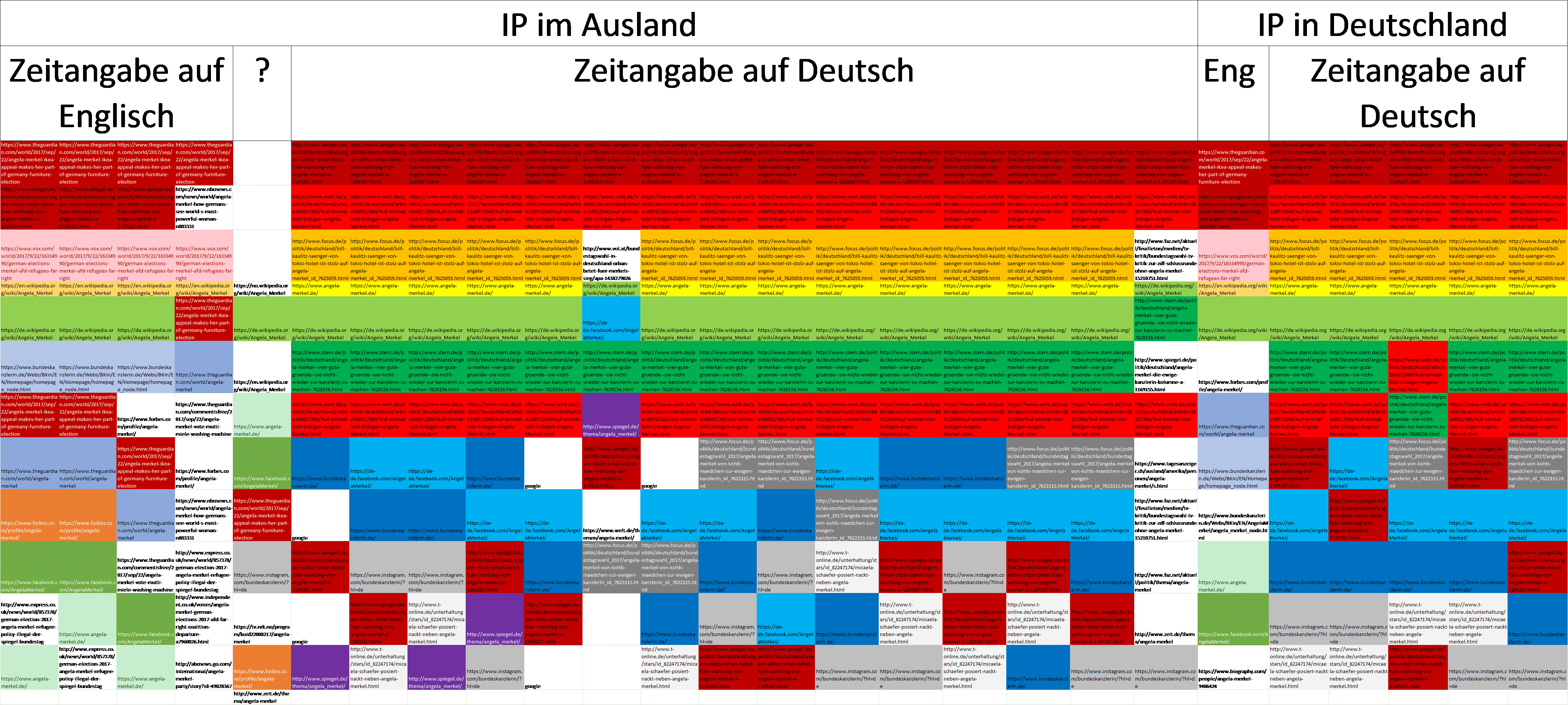}
    \caption{Search result lists on September 22, 2017 at 4pm for the search term "Angela Merkel". They are sorted into result lists from users with an IP address outside of Germany (6x Switzerland, 3x Austria, 2x Belgium, 2x USA, 2x France, 2x Great Britain, Denmark, Norway) separated into columns. The last 6 columns contain users with an IP address based in Kaiserslautern. If top stories were present, the publication time ("vor 4 Stunden", "1 hour ago") indicated which language was assigned to the user by Google. The deviant clusters were accompanied by English publication times, even with the user from Kaiserslautern. The result lists in the cluster themselves are very similar to each other.}
\end{figure*}

Even though the patterns in the deviant clusters remain complex and also due to the relatively low number of users with constantly German websites, a distinction between "German" and "non-German" users is clearly visible. It is unclear which aspects factor into Google's decision whether a person is German-speaking and should only receive German results. Users who also received English results showed very similar result lists, regardless of their location (in this case Austria, Belgium, Great Britain and Denmark). One exception to this is a person from Norway, who did not receive any top stories -- therefore it is unclear which language Google has assigned to them. It is however noteworthy that this user received two Norwegian websites and has received top stories for another search term ("Alexander Gauland") at the same search time, which had Norwegian publication times. Another exception is one of the Swiss users, who received a lot of results not present in the others' result lists.\\
This results in a not completely isolated, but noticeably different media space for the users receiving English (or other foreign) results while living in Germany (or at least using a router located there).
\subsection{Possible explanations}
In the end the simplest explanation is probably the correct one: The type of deviant search result lists for German IP locations can easily be recreated by logging into one's Google account and changing the search language settings. To accomplish that, one must go to the search settings on the Google website and change the language settings to the desired language, e.g. French. Another visit to the Google website shows that the navigational buttons are now displayed in the selected language. If the term searched for results in the display of top stories, these will be accompanied by the publications times in the selected language. Additionally the organic results contain some deviant links identical to the one we found when searching for the search terms used in this project: Links to English Wikipedia pages, confusing CSU, CDU, AfD and SPD with identical acronyms, more common in the selected language (e.g. "Colorado State University", "Charles Darwin University", "Agence Fran{\c{c}}aise de D{\'e}veloppement", "symphysis pubis dysfunction"), and also German websites that indeed belong to the German political party or their respective social media profiles. The similarities to the deviant results found within our data set are strong -- but this does not necessarily proof that the search language settings are the only factor resulting in these deviant result lists. We follow the principle of "Occam's razor", which states that the simplest solution tends to be the right one. We therefore conclude that the deviant result lists originate in the personal settings of the respective data donor. The result is a combination of language filter, regionalization (most likely depending on the IP location) and the most relevant websites for the given search term.\\
It is important to note that the cases shown here did not contain news about different political perspectives on the politicians or the political parties. According to Eli Pariser's dreaded content-specific filter bubble, wherein some users receive a drastically different political perspective than others could not be found in our manually inspected cases. English news came from The Guardian, French news from Le Figaro, the Russian websites were Wikipedia or Wikimedia pages. In a lot of cases the search results of a "foreign" user differed from specific aggregate topic pages, like http://www.faz.net/aktuell/\\politik/thema/martin-schulz compared to the result lists of more regionalized "German-speaking" users.\\
It still brings up the question what kind of information people belonging to one of the minorities of a country, like the first and second generation of migrants, should receive and from which source, so that these discussion forums overlap as far as possible.

\section{Summary}
We come to the conclusion that, since almost all of the search result lists show sufficient overlap -- regardless of content -- the algorithmically based creation and reinforcement of isolated filter bubbles is not present. It is noteworthy that even though users who received foreign websites from Google (probably due to different search language settings of these users) showed some overlap in their results, their deviant clusters can be a sign of a filter bubble. Though this is unavoidable due to the difference in languages, this means that, based on the four mechanisms that have to be working together to create these dangerous filter bubbles, one must still inspect the contents. Manual inspection and a number of received top-level domains mostly display mainstream media, but there are also stray top-level domains and URLs that are harder to classify. This would require interdisciplinary analysis of the data and preferably further research.\\
But what about the question whether the algorithm provides all of us with one-sided content? Firstly the large amount of received top-level domains contradicts this, but that does not answer the question if there could have been other top-level domains that should have been displayed, or if some top-level domains appeared more or less frequently. What could be used to make such a comparison? One possible dimension would be to use the popularity of the top-level domains according to all internet users, as is continuously measured by the Nielsen-Online-Panel. The Nielsen-Panel is based on a representative user group, whose internet behavior is measured in great detail. For every top-level domain there is a score called the "active reach" based on the measurements of the panel that shows the percentage of users that visited that top-level domain at least once in a given time period\footnote{This score is not very meaningful for extremely low scores, e.g. less than a hand full of people visited this website. This will be noted in the Nielsen Report.}.\\

\begin{figure}[ht!]
    \centering
    \includegraphics[width=0.5\textwidth]{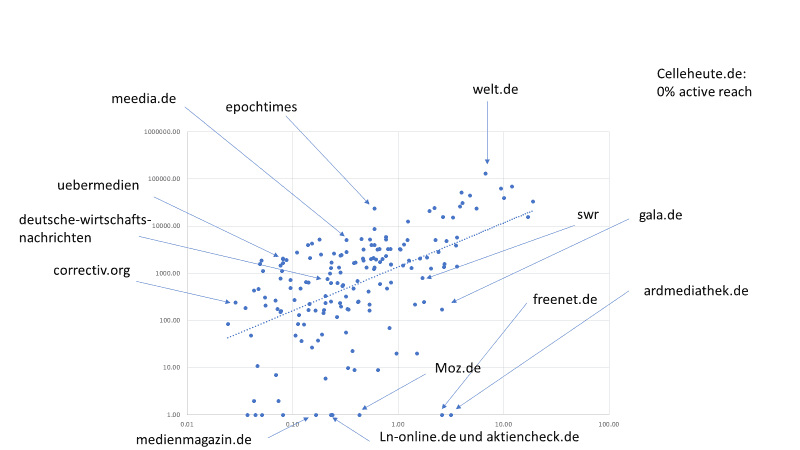}
    \caption{The diagram shows the "active reach" (percentage of users that visited the top-level domain at least once) for the different top-level domains according to the August edition of the Nielsen Report and the total number of search results from August 2017 of our data set. It is a double logarithmic application, where some sources with strong deviation from the trend line have been marked.}
\end{figure}

This leads to another chicken-and-egg problem, since Google is definitely the most prominent search engine in Germany and a great number of visits to a website are done via referral by Google. This means that there is an unknown proportion of visits that only occurred because of Google and hence the numbers do not represent a popularity measure independent of Google's search engine algorithm. Therefore, the following findings are more indicative of further research questions than the results of the \#Datenspende project.\\
Figure 34 shows the application of the "active reach" of a domain against the URLs of the top stories (not the organic searches) of this domain supplied by Google to the data donors in the same period. Since the differences per axis are very large (from 0.01 to 100\% or from 1 to 1,000,000 URLs displayed), a double logarithmic application was chosen. The pattern of the data allows to create a linear regression by this application. The formula y = 1,373 * active reach\textsuperscript{0.9} essentially means that the number of URLs delivered correlates almost linearly with the range, with a small damper for longer ranges.\\
In the Figure, some points are picked out which deviate from this pattern by a factor of approximately 5 upwards or downwards. This means that they are delivered at least 5 times more or 5 times less than expected according to the formula. On the one hand, this includes media that report on media such as correctiv.org, meedia.de and uebermedien. Correctiv.org is a correction website that takes up and corrects incorrect messages. While this top-level domain in the Nielsen Panel has no reach in August, one of its URLs will be delivered to 245 people in the data donation project, almost five times more than expected. Nevertheless, this is less remarkable than first thought, as it concerns exactly one top story, which is delivered to all donors at this search time (August 24, 2017, 12pm, search term "Alexander Gauland"). Since we have already found that the search result pages are very uniform overall, such a decision always affects everyone at the time of the search. This is a one-time event that suddenly raises the number of URLs delivered above the expected value. Articles by meedia and uebermedien are delivered 10 and 16 times more than expected. It is important to emphasize once again that the comparison with the Nielsen reach is methodologically difficult: neither does the "active reach" represent a popularity measure independent of Google's search engine, nor can popularity be regarded without restriction as a direct measure of the content quality of the contributions of the top-level domains.\\
It may still be an interesting observation that the following 10 sources are the most overrepresented (regarding the expected value from the regression):
\begin{enumerate}
\item	Epochtimes (with 240.004 links in all search result lists, expected were 840)
\item	Chiemgau24.de (1.905 links, expected 22)
\item	Jungewelt.de (1.595 links, expected 81)
\item	Stimme.de (5.259 links, expected 274)
\item	Ka-news.de (4.047 links, expected 218)
\item	Tichyseinblick.de (4302 links, expected 235)
\item	Welt.de (131.793 links, expected 8120)
\item	{\"U}bermedien.de (2102 links, expected 131)
\item	Cicero.de (2785 links, expected 174)
\item	Butenunbinnen.de (2058 links, expected 131)
\end{enumerate}
As with all relative scores, the most outliers are sources that did not appear often in the search results. The media with the highest reach are however still represented within the top stories delivered by Google. The combination of the websites of persons and political parties, the Wikipedia entries and news websites, all found in the organic search results, shows a broad view of politicians and parties, which on the whole allows for comprehensive information and opinion-forming. It should also be mentioned that further combining the search terms used in this project would allow for a more detailed search, in the case of an unsatisfactory first search result list.
\subsection{Generalization of the results: possibilities and boundaries}
The results shown in this report can not be applied to all Google users in Germany, since the data donors are not a representative sample (caused by their own decision to participate). Interestingly, the donors who seem to have used a different search language could invalidate the argument that the set of users was so homogeneous that this alone explains the low degree of personalization observed. It appears improbable that all English-speaking users from different countries are homogeneous in personalization, yet Figure 33 shows remarkable similarities between their search result lists. Here too, we follow the principle of Occam's razor and assume that the search result lists are mainly determined by search language settings, the location (e.g. based on the IP address) and the general relevance of the website, followed by previously visited websites to a lesser extent and possibly some personalized links. Therefore it would surprise us if further studies show highly deviant degrees of personalization.\\
In every case, our observations are only valid for the time period and search terms used in this project. We can not preclude the possibility of the degree of personalization changing over time, but this study shows how this can be monitored simply, cost-efficiently, flexibly (regarding the search terms) and automatically. This holds true for every search engine.
\subsection{Demand for suitable interfaces to examine the filter bubble theory in social networks and other intermediaries}
This does not apply to social networks like Facebook, other social media like Twitter, Instagram or YouTube. At the moment it would only be possible with great effort to filter out and centrally collect all of a user's political content in the news feed of a Facebook account, for example. ProPublica has tried such a project during the federal election of 2017. An API that allows for selective access to these contents does not exist, so the only choice would be to require the users to allow complete access (privacy concerns) or to ask them to submit screenshots of political content. The latter is neither easy, nor can the data be analyzed automatically, since the content would have to be described manually to allow for searching and summarizing the data.\\
Hence a demand for the necessary interfaces arises, to investigate all social media and social networks, whose degree of personalization could lead to algorithmic creation or reinforcement of filter bubbles. A suitable control of the degree of personalization relieves the respective companies -- as long as the degree remains low -- of the demands of the society to gain further insight into their software code. If a high degree of personalization (of relevant content) and a certain degree of isolation of the filter bubbles becomes evident, further investigations are required to determine to what extent the content has extremely different perspectives and makes social discourse more difficult. Such a staggered model of the depth of insight through black box analysis depending on the measured degree of personalization would thus help both sides.

\section{Acknowledgment}
We would like to take this opportunity to thank Landesmedienanstalten Bayern (BLM), Berlin-Brandenburg (mabb), Hessen (LPR Hessen), Rheinland-Pfalz (LMK), Saarland (LMS) and Sachsen (SLM) who supported the project, as well as Spiegel Online as our media partner. We would also like to express our gratitude towards Uwe Conradt, director of the LMS, who suggested this project to us. We also thank Dr. Anja Zimmer, director of the mabb, and Siegfried Schneider, president of the BLM, who -- together with Uwe Conradt -- were particularly committed to the project. Our special thanks go to Adrian Gerlitsch (BLM) who supervised the project and has always and tirelessly found constructive solutions to all problems.

\section{Appendix}
\textbf{A) Category System}\\
The category system is used to assign domains that were displayed as part of the data donation. It was developed by the Bayerische Landeszentrale f{\"u}r neue Medien (BLM) who also carried it out.
\subsection{Main categories}
Every domain is assigned to one of the main categories. In cases of conflict the most concise assignment is chosen.
\begin{enumerate}
\item[a] \textbf{Owned Content}: The category Owned Content contains all channels that allow the parties an unfiltered display of opinion, either through a personal website of a member of the party, a local branch, or the party itself. Social media channels also belong to this category, but will be assigned to their own category in this categorization scheme.
\item[b] \textbf{Social Media}: Social Media describes all channels that focus on the social interactions between provider and users. The domains allow for networking, i.e. to exchange ideas with each other and to create and relay media content for individuals, specific groups, or the public. Examples for networks that belong to this category are Facebook, Twitter or Instagram.
\item[c] \textbf{Wikipedia}: All Wikipedia domains are assigned to this category.
\item[d] \textbf{Media}: Domains that are operated by a media provider are assigned to this category. A website is considered a media provider if it provides journalistic content and acts as the online offshoot of classic print media, a private television or radio provider, or a public-service broadcaster. Foreign journalistic content as well as online-only providers also belong to this category (see sub-categories). Examples are: www.schwarzwaelder-bote.de, news.sky.com, www.news.de.
\item[e] \textbf{Freemail portals}: Freemail portals offer free email addresses and a mailbox to send and receive emails. Most of these websites also contain news columns. The most common Freemail portals belonging to this category are web.de, yahoo.com and gmx.net. Freemail domains that appear in the results of Google News will not be assigned to any sub-categories, only the main category.
\item[f] \textbf{Publicly funded}: The category "publicly funded" contains all domains financed by public funds, e.g. the Federal Agency for Civic Education (Bundeszentrale f{\"u}r politische Bildung (bpb)) and the online presence of cities and municipalities.
\item[g] \textbf{Other}: Every domain that does not explicitly belong to any of the before-mentioned categories will the assigned to the this category.
\end{enumerate}
\subsection{Sub-categories}
Sub-categories contain domains that were previously assigned to the "Media" category. The other main categories are not specialized any further. Every domain can only be assigned to one sub-category.
\begin{enumerate}
\item[a] \textbf{Print}: The category contains all online offerings from classic German print media (newspapers and magazines). Domains are included if they originate in Germany and publish German content. Example include: spiegel.de, faz.net, welt.de.
\item[b] \textbf{TV}: All online offerings of approved German private television providers are assigned to this category. Both origin and published language of the content is German. Example include: rtl.de, pro-sieben.de
\item[c] \textbf{Public service providers ({\"O}RR)}: This category contains all online offerings of German public-service broadcasters, like their online branches for radio and television channels. This category can contain daserste.de, ndr.de, zdf.de.
\item[d] \textbf{Online Only}: Media offerings who publish their content exclusively on the internet are assigned to this category. Examples include: 02elf.net, www.promiflash.de.
\end{enumerate}

\bibliographystyle{ACM-Reference-Format}
\bibliography{Report_What_did_you_see}
\nocite{*}
\end{document}